\documentclass[conference]{IEEEtran}
\IEEEoverridecommandlockouts
\usepackage[T1]{fontenc}
\usepackage[utf8]{inputenc}
\usepackage{cite}
\usepackage{amsmath,amssymb,amsfonts}
\usepackage{algorithm}
\usepackage{algpseudocode}
\usepackage{graphicx}
\usepackage{textcomp}
\usepackage{listings}
\usepackage{float}
\usepackage{booktabs}
\usepackage{tabularx}
\usepackage{url}
\usepackage{xcolor}
\usepackage[hidelinks]{hyperref}

\hypersetup{
  unicode=true,
  pdftitle={A Modular, Topology-Aware Software Stack for Entanglement-Based Distributed Quantum Computing},
  pdfauthor={Luke Andreesen, Shobhit Gupta, Sean Sullivan, Manish Kumar Singh},
  pdfsubject={Distributed quantum computing},
  pdfkeywords={distributed quantum computing, quantum circuit compilation, quantum circuit partitioning, quantum networking, quantum scheduling, hardware-software co-design},
  pdfdisplaydoctitle=true
}

\makeatletter
\def\ps@preprint{%
  \def\@oddhead{\normalfont\small\hfil\thepage}%
  \def\@evenhead{\normalfont\small\hfil\thepage}%
  \def\@oddfoot{}%
  \def\@evenfoot{}%
}
\makeatother

\newif\ifrevisionmode\revisionmodefalse

\graphicspath{{figures/}}

\lstdefinelanguage{OpenQASM}{
  morekeywords={OPENQASM,include,qreg,creg,qubit,bit,gate,
    measure,reset,barrier,if,for,in,let,def,box,gphase},
  morekeywords=[2]{h,x,y,z,s,t,rz,rx,ry,cx,cz,cp,swap,u,
    catent,catdisent,rcx,rcz,rcp,rcry,rswap},
  sensitive=true,
  morecomment=[l]{//},
  morestring=[b]",
}
\lstdefinestyle{qasm}{
  language=OpenQASM,
  basicstyle=\footnotesize\ttfamily,
  keywordstyle=\color{blue!70!black}\bfseries,
  keywordstyle=[2]\color{purple!70!black},
  commentstyle=\color{gray}\itshape,
  stringstyle=\color{teal},
  numbers=left, numberstyle=\tiny\color{gray}, numbersep=5pt,
  frame=single, framerule=0.3pt, rulecolor=\color{black!30},
  showstringspaces=false,
  columns=fullflexible, keepspaces=true,
  breaklines=true, xleftmargin=1.8em,
}

\def\BibTeX{{\rm B\kern-.05em{\sc i\kern-.025em b}\kern-.08em
    T\kern-.1667em\lower.7ex\hbox{E}\kern-.125emX}}

\begin{document}

\bstctlcite{IEEEexample:BSTcontrol}

\title{A Modular, Topology-Aware Software Stack for Entanglement-Based Distributed Quantum Computing}
\author{
\IEEEauthorblockN{Luke Andreesen, Shobhit Gupta, Sean Sullivan, and Manish Kumar Singh}
\IEEEauthorblockA{memQ Inc., Chicago, IL 60615, USA}
}

\maketitle
\pagestyle{preprint}
\thispagestyle{preprint}

\begin{abstract}
Distributed quantum computing (DQC) seeks to scale beyond the limits of monolithic processors by interconnecting multiple quantum processing units (QPUs) through entanglement-based links. Realizing this vision requires the co-design of hardware and software across the domains of quantum networking, compilation, and scheduling, yet existing tools remain fragmented between monolithic circuit compilers and long-distance quantum network simulators. We present an open-source,  topology-informed framework for the compilation and scheduling of distributed quantum programs. Given an input circuit and a description of the target network, the framework partitions and reconstructs the circuit into a distributed program that respects the specified inter- and intra-QPU topology, covers cross-QPU operations through gate and state teleportation, and schedules the result under either deterministic or stochastic entanglement-generation models. By accepting and emitting standard OpenQASM, the framework interoperates with existing monolithic toolchains, and its standardized module interfaces allow partitioning and scheduling strategies to be interchanged and benchmarked. Using this framework, we show that the best-performing compilation strategy varies across the tested circuits and network configurations, and that both network topology and intra-QPU connectivity substantially affect the entanglement cost of execution. These findings underscore the need for a co-design approach to distributed quantum computing, which our framework is designed to support.
\end{abstract}


\section{Introduction}

Distributed quantum computing (DQC) offers a path toward scaling beyond the limitations of monolithic processors. These limitations stem from the difficulty of scaling qubit count, connectivity, and control within a single device; distributing a computation across several interconnected processors offers a promising route around them. In addition to local operations of a monolithic processor, a distributed quantum processor will utilize entanglement-based quantum links to orchestrate cross-QPU operations. Thus, DQC combines the entanglement distribution and resource management of quantum networking with network-aware circuit compilation and scheduling, layered atop the local operations of a monolithic processor.

These functional attributes present a large design space in terms of optimal local and remote protocols, inter- and intra-QPU compilation algorithms, remote entanglement resource management and routing strategies, quantum interconnect hardware specifications, quantum network topologies, and the choice of error-correcting codes and decoding strategies. Therefore, to realize the promise of DQC, we need to adopt a hardware-architecture-software-algorithm co-design approach, where low-level hardware details are mapped through the stack to the high-level algorithm. The performance characteristics of underlying hardware, including qubit lifetimes, gate fidelity, intra-QPU connectivity, quantum-link rate and fidelity, and inter-QPU connectivity, together determine the optimal architecture of the distributed QPU. The software layer must account for these hardware-level granular details to decide the optimal allocation of resources for the execution of the algorithm, including qubit allocation and cross-QPU operation orchestration.

An ideal software framework for DQC must extend existing monolithic software frameworks with several domain-specific layers to accurately simulate and orchestrate entanglement-based distributed quantum computing. These layers include a network-aware circuit compiler that performs circuit partitioning and qubit allocation across QPUs, and a scheduler that dynamically orchestrates intra- and inter-QPU operations, powered by a discrete-event simulator that produces heralded `clicks'---signals that indicate successful remote entanglement generation. 

In this work, we present an open-source framework for the hardware-informed co-design of distributed quantum computing. Here, co-design refers to the joint selection of compilation and scheduling strategies, inter- and intra-QPU network topology, and hardware specifications, rather than optimizing any one of these in isolation. The framework exposes a user-friendly front-end interface through which users can specify hardware properties and network topology and invoke the compiler and scheduler, without engaging with the low-level implementation of each layer. The framework is intended to bridge the gap between hardware and algorithms and to support the co-design of distributed quantum processors, algorithms, and network protocols.

The remainder of this paper is organized as follows. Section~\ref{sec:background} provides background on the primitives that underlie entanglement-based distributed quantum computing and reviews related software frameworks. Section~\ref{sec:architecture} presents the architecture of the framework, describing its functional modules and their input and output specifications. Section~\ref{sec:methods} details the compilation and scheduling methods implemented within these modules, including circuit preprocessing, the partitioning algorithms, circuit reconstruction, and schedule generation. Section~\ref{sec:results} presents a series of demonstrations which emphasize the importance of granular control over the software's inputs and strategies. Finally, Section~\ref{sec:discussion} discusses the broader implications of these results and Section~\ref{sec:futurework} outlines directions for future work.



\section{Background \& Related Work}
\label{sec:background}

\subsection{Distributed Quantum Computing Primitives}

\begin{figure*}[t!]
    \centering
    \includegraphics[width=0.8\textwidth,height=0.5\textheight,keepaspectratio]{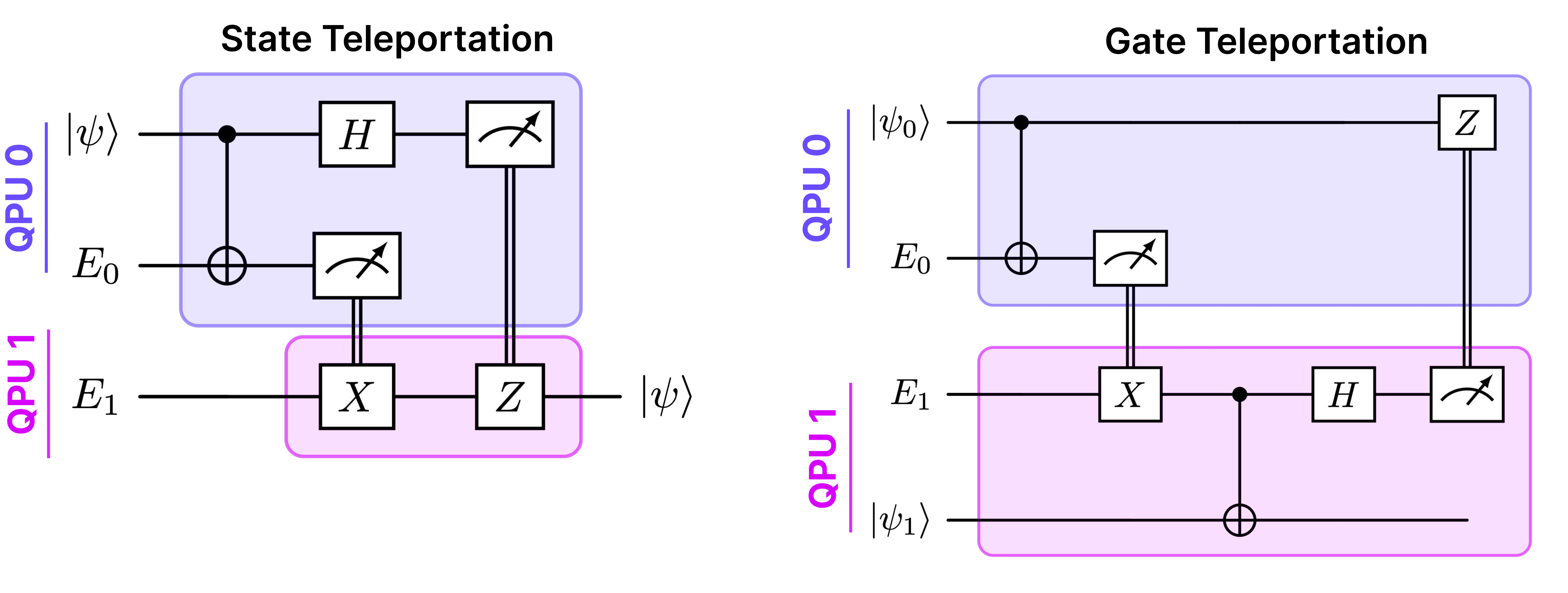}
    \caption{State Teleportation (left) and Gate Teleportation (right) circuits. The state teleportation circuit transfers an arbitrary qubit state from one QPU to another QPU using an EPR pair. The gate teleportation circuit allows a two-qubit gate to be executed between two qubits sitting on distinct, interconnected processors.} \label{fig:teleport}
\end{figure*}

Entanglement-based distributed quantum computing relies on the use of quantum interconnects between processors. The interconnects provide entangled Einstein--Podolsky--Rosen (EPR) pairs, which are used as a resource to enable quantum state teleportation and quantum gate teleportation. The state teleportation protocol \cite{bennett1993, bouwmeester1997}, as shown in Fig. \ref{fig:teleport}, is used to transfer an arbitrary qubit state from one QPU to another QPU by using an EPR pair. Gate teleportation, pictured in Fig. \ref{fig:teleport}, is a protocol which allows a two-qubit gate to be executed between two qubits sitting on distinct, interconnected processors \cite{eisert2000, Yimsiriwattana2004}.

A primary metric considered in this work is the number of EPR pairs consumed, since generating these pairs is far more time-intensive than executing local gates. An optimal compiler supports both protocols, selecting between them to minimize EPR consumption, as each excels in a distinct regime. When a qubit interacts with many others scattered across different QPUs, remote gates are more efficient, avoiding repeated shuttling of qubits between processors. Conversely, state teleportation is preferable when a qubit undergoes many subsequent operations with qubits on a single destination QPU, since one move makes all of those interactions local.

\begin{figure}
    \centering
    \includegraphics[width=0.8\linewidth]{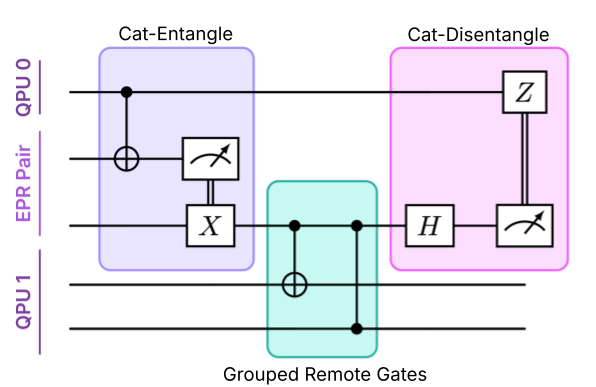}
    \caption{{Circuit-level implementation of the Cat-Entangler and Cat-Disentangler protocols. The Cat-Entangler consumes a single shared EPR pair to distribute the state of a control qubit onto a communication qubit on a remote QPU, allowing one or more gates controlled by that qubit to be executed remotely.}} \label{fig:circuit-catent}
\end{figure}

The framework models remote controlled gates using the `Cat-Entangler' and `Cat-Disentangler' protocols \cite{Yimsiriwattana2004}, shown in Fig.~\ref{fig:circuit-catent}. The Cat-Entangler protocol allows two QPUs to share a control qubit using a single entangled pair, so that a gate controlled by that qubit can be executed on the remote QPU. Once the required remote gates have been applied, the Cat-Disentangler protocol disentangles the control qubit from the target QPU, returning it to its original state.

\subsection{Related Work}
Existing software frameworks lack the functionality to enable end-to-end co-design of a distributed QPU. On one hand, there are several software frameworks tailored to monolithic processors, including circuit compilers \cite{superstaq2023}, transpilers \cite{qbraid2026}, and general-purpose circuit simulators such as Qiskit Aer \cite{qiskit2024} and PennyLane \cite{pennylane2022}. On the other hand, the long-distance quantum networking community has developed several software frameworks and discrete-event simulators such as SeQUeNCe \cite{Sequence2021}, NetSquid \cite{NetSquid2021}, and QuISP \cite{QuISP2022}.

A substantial body of work on circuit partitioning and distributed quantum compilation has been proposed, wherein a monolithic circuit is partitioned into subcircuits for execution across a network of interconnected quantum computers \cite{beals2013, Zomorodi2018, Martinez2019, dadkhah2021}. Although this body of work provides a strong foundation for distributed quantum compiler algorithms and approaches, a frequent limitation is reliance on idealized hardware models, particularly optimistic QPU-connectivity and network-topology assumptions. Additionally, while prior work provides techniques which perform strongly for certain quantum circuits, they often fail to generalize this performance across a range of representative circuits. While hardware-focused approaches to this problem have been explored \cite{du2025, AndresMartinez2024}, an end-to-end approach which fully enables the co-design of a distributed quantum processor, from hardware specifications to quantum algorithm choice, remains an open problem.

Table~\ref{tab:software-comparison} in the appendix provides a detailed comparison of selected distributed quantum computing software. SeQUeNCe, NetSquid, and QuISP provide network-level simulation and protocols, while Qoala leverages NetSquid to provide application and execution layers for quantum-network simulation. Among the distributed compilers considered---Pytket-DQC, Ferrari et al.~\cite{ferrari2023}, Mengoni et al.~\cite{mengoni2026}, and Kaur et al.---only Pytket-DQC is open source, and none produces a time-resolved operation schedule. Additionally, most of these existing frameworks do not provide full control over both intra- and inter-QPU connectivity. Our open-source framework provides both distributed-circuit compilation and timestamped operation scheduling across arbitrary QPU topologies.

The primary goal of this framework is to address two key resource considerations in the design of DQC systems: the number of EPR pairs consumed by a distributed program and the time required to execute it. This software stack is not intended to serve as a comprehensive quantum network simulator and currently does not model fidelity or noise, as simulators such as SeQUeNCe and NetSquid do. Additionally, the stack does not yet account for decoherence, entanglement distillation, or other physical phenomena fundamental to distributed quantum computing; support for such features is left for future work. We do not propose an entirely novel partitioning algorithm. Instead, we provide a framework that integrates multiple strategies adapted from prior work and supports the streamlined development and evaluation of new partitioning algorithms. Finally, the framework targets NISQ-era distributed quantum computing and does not support logical-level compilation or quantum error correction.
\section{Software Architecture}
\label{sec:architecture}

\begin{figure}[t]
    \centering
    \includegraphics[width=1\linewidth]{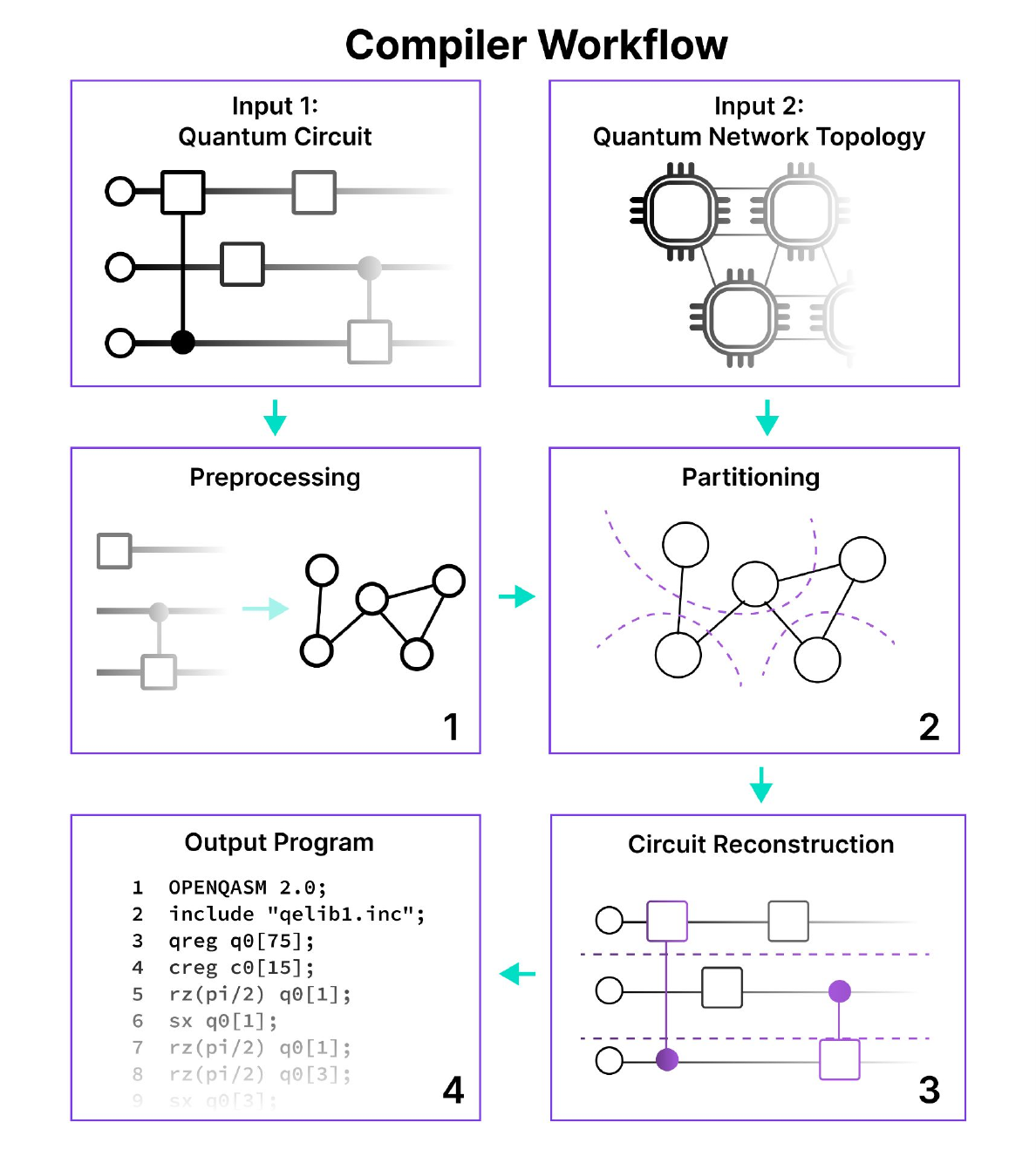}
    \caption{Overview of the key steps in the compiler workflow. The compiler takes two inputs: (i)~a quantum circuit to be executed and (ii)~a quantum network topology describing the cluster of interconnected QPUs. In the preprocessing stage~(1), the circuit is converted into a graph-based representation that captures both circuit dynamics and network topology. The resulting interaction graph is then partitioned~(2) across the available QPUs, with cut edges (dashed) denoting interactions that span partition boundaries and therefore require inter-QPU communication. During circuit reconstruction~(3), the partitioned assignment is mapped back onto circuit form, inserting the operations needed to realize cross-partition gates (highlighted) consistent with the network topology. Finally, the compiler emits an output program~(4) as an OpenQASM file targeting the distributed architecture, which can be paired with a time-resolved execution schedule for the cluster.} \label{fig:xdqc_workflow}
\end{figure}

Our toolchain transforms a standard quantum circuit into a program executable across a network of interconnected QPUs, and then generates an execution schedule. The compiler partitions the circuit's qubits and operations across the available QPUs, inserting the remote operations required to span partition boundaries, and emits a distributed circuit in OpenQASM format. This compiled circuit serves as input to the scheduler, which assigns both local and network operations to discrete time slots. An overview of this workflow is shown in Fig.~\ref{fig:xdqc_workflow}.

\subsection{Quantum Network Constructor}
The key focus of this distributed quantum computing toolchain is configurability and modularity. The optimal configuration of quantum networks, spanning from datacenter use cases to long-range networking, remains an open question. Thus, the ability to easily configure software to varying hardware configurations is crucial. Our software provides hardware-specific compilation and scheduling. Given a network configuration, including exact mappings of intra- and inter-QPU connectivity, the compiler produces a program tailored to that topology. The software ensures that all local and remote operations are physically feasible.

\begin{figure}
    \centering
    \includegraphics[width=.8\linewidth]{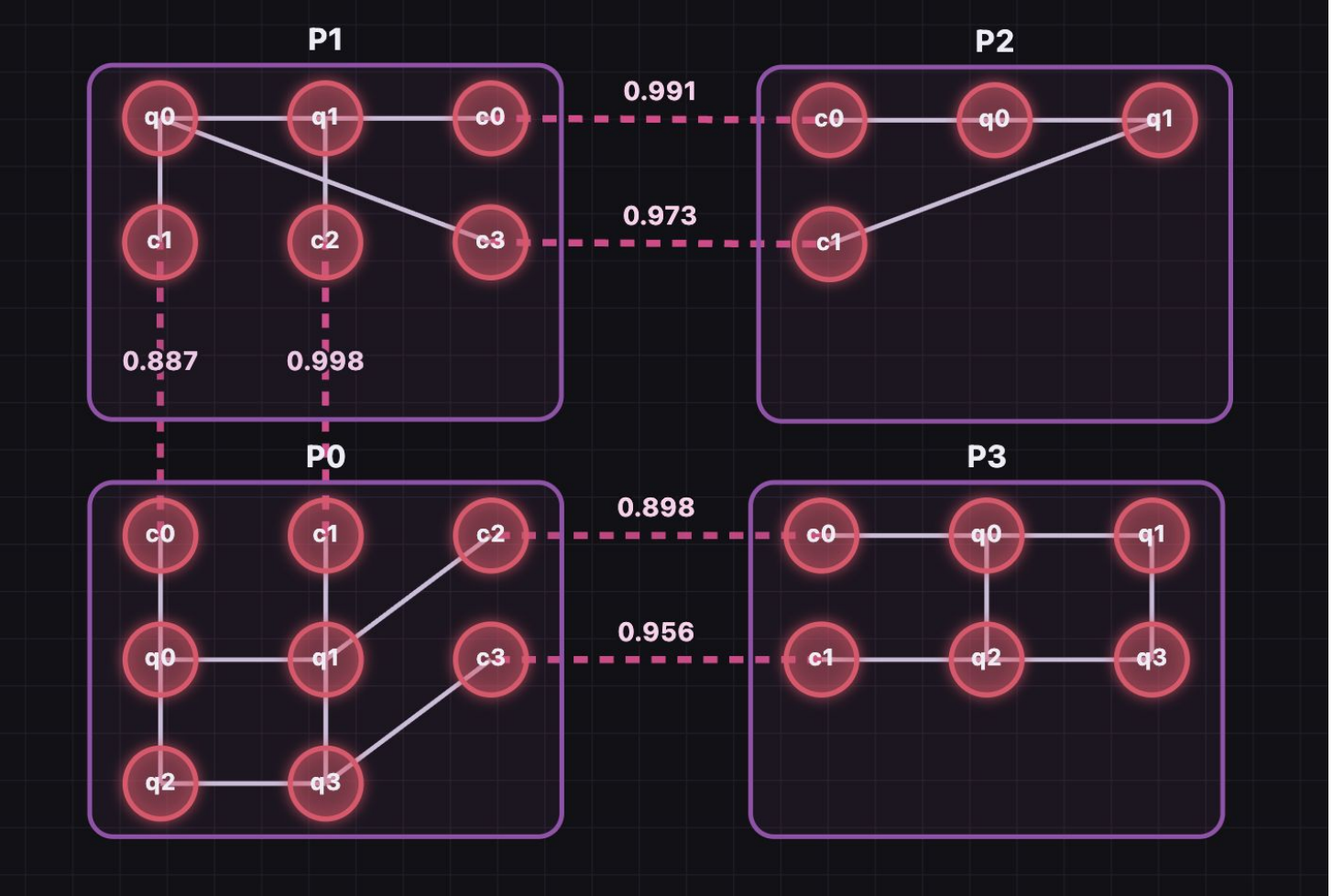}
    \caption{A sample network configuration generated via Quantum Network Constructor. This illustrative system shows four QPUs, with varying interconnect fidelities (dotted lines). Solid lines indicate intra-QPU connections. Circles with `q' labels, e.g. `q0', represent computation qubits, while circles with `c' labels, e.g. `c0', represent communication qubits.} \label{fig:network_config}
\end{figure}

The network configuration is passed to the compiler in a standard JSON file format. To streamline the network specification process, thus enabling rapid experimentation with varying topologies, we provide the Quantum Network Constructor (QNC), a graphical tool which allows users to easily describe a particular network. As shown in Fig.~\ref{fig:network_config}, the QNC is used to specify local connectivity within each QPU, remote connectivity between communication qubits of each QPU, and link fidelity. Importantly, networks can be constructed with varying intra-QPU topologies and qubit counts, allowing arbitrary network configurations to represent any desired QPU combination. Additionally, a module of only communication qubits can be constructed using the QNC, effectively acting as a quantum router. Communication qubits are automatically created when inter-QPU links are formed between qubits of any pair of processors.

After configuring the network, the QNC is used to automatically generate a JSON file representation of this network. The standard file format is then used as an input to the compiler stack, which internally converts this file to a graph representation, which is then utilized in the compilation process.

\subsection{Compiler}
The key to executing distributed quantum circuits is the compilation process, in which a standard quantum circuit, represented in OpenQASM or as a Qiskit circuit, is converted to a circuit which can be executed on a network of interconnected quantum computers. The core of compiling is to partition the qubits and operations of a circuit into subcircuits, which are then allocated to and executed on individual QPUs within the network.

While the algorithms used in the compilation process are described in depth in Section~\ref{sec:methods}, the general process is as follows: first, the input circuit is mapped to a graph representation. Next, this graph is partitioned while obeying the input network topology, where the number and sizes of the partitions correspond to the number of QPUs and their respective capacities. This mapping can be dynamic in time, meaning qubits move between QPUs during the execution of the circuit. To resolve this dynamic mapping, the compiler inserts state teleportation primitives. The compiler then resolves all two-qubit gates between qubits mapped to distinct QPUs by inserting remote gate primitives. The resulting circuit is output in OpenQASM format, with custom gates inserted to represent quantum network operations.

\subsection{Scheduler}
The final core module of this framework is a distributed quantum circuit scheduler. The purpose of the scheduler is to assign timestamps to each of the operations in the compiled distributed program. Specifically, the scheduler uses as input a directed acyclic graph (DAG) representation of the distributed program, which captures all inter-operation dependencies. The scheduler strictly enforces these circuit dependencies while giving start times to local operations (single-qubit gates, multi-qubit gates, and measurements) as well as remote operations, which include entanglement generation requests, remote gates, and state teleportation. 

Due to the stochasticity of the entanglement generation process, it is important for the scheduling process to be dynamic. This work provides a discrete-event scheduler, which simulates stochastic heralded photon-based entanglement generation in the scheduling process. Specifically, upon a request for an EPR pair resource between two QPUs, a stochastic process based on entanglement generation rates for particular protocols is used to generate a `click', after which a remote operation begins using the EPR pair. This work also provides a deterministic scheduling mode, in which all operations, including entanglement generation, take a constant amount of time.
In either case, the scheduler emits a fully timed sequence of local and network operations that respects all circuit dependencies. Because this schedule specifies exactly when each entanglement attempt, remote gate, and local operation must occur, it provides a clear path towards execution of distributed quantum programs on actual hardware.

\begin{figure*}[!t]
    \centering
    \includegraphics[width=\textwidth]{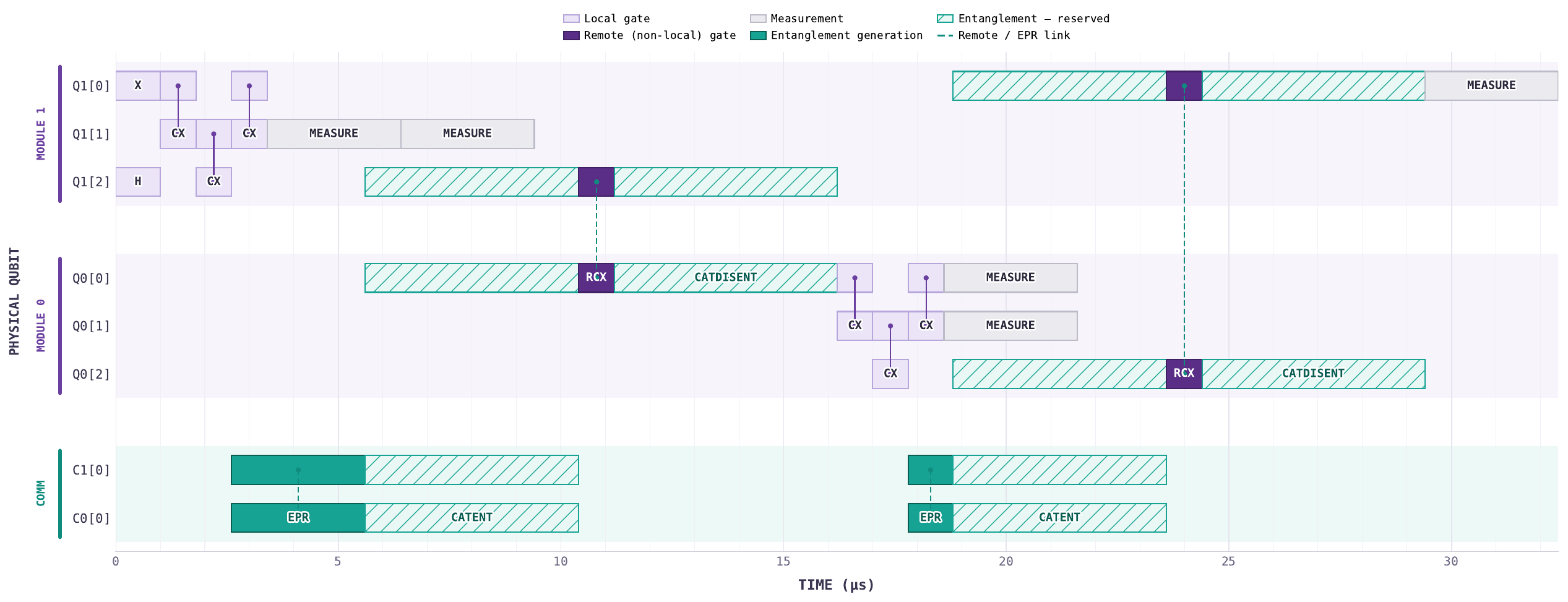}
    \caption{\textbf{Example distributed circuit schedule.} Produced and displayed as a Gantt chart by the scheduler module. Note that EPR generation times are shown at a compressed scale for visual clarity, as they would otherwise dominate the schedule and obscure the gate operations.} \label{fig:gantt}
\end{figure*}

The output of the scheduler is an object which assigns each local operation (local single- and multi-qubit gates, measurements) and remote operation (entanglement generation, remote gates, state teleportation) to a specific start and end time, with the duration informed by the specified hardware parameters. This object can then be serialized to a JSON format, or a visual representation can be produced by the software to display the schedule as a Gantt chart, as shown in Fig.~\ref{fig:gantt}.

\section{Compilation \& Scheduling Methods}
\label{sec:methods}
\subsection{Circuit Preprocessing}
In order to create a modular software architecture suitable for research, this toolchain first conducts a preprocessing step on the input circuit to ensure all partitioning algorithms receive a standardized input. The input to this step is a standard OpenQASM circuit~\cite{openqasm3}. The program's abstract syntax tree (AST) is first parsed and normalized to standard conventions for gate and register declarations. This normalized AST is then converted to a DAG, which serves as the intermediate representation used throughout the compilation process. In the DAG, each node represents a quantum operation, and directed edges between nodes represent data dependencies. This DAG representation is used to decompose the circuit into a sequence of layers, where each layer is a set of circuit operations which can be executed concurrently upon completion of the previous layer.

This preprocessing step is what allows the compiler to support the diverse ecosystem of high-level quantum programming languages. Because OpenQASM is a widely used, open-source format, any program that can be expressed in or converted to OpenQASM is supported automatically. Qiskit \cite{qiskit2024}, for example, can export its circuits directly to OpenQASM, while transpilation tools such as the qBraid Transpiler \cite{qbraid2026} can translate arbitrary source representations into OpenQASM. 

\subsection{Partitioning}
At the core of the distributed compilation process is the partitioning step, which takes two standardized inputs: the DAG representation of the circuit produced by the preprocessing step, and the network configuration, provided as a JSON file. From these, the partitioner produces a standardized output describing the location of every circuit qubit throughout execution. Concretely, each partitioning algorithm outputs a matrix whose rows correspond to the qubits of the input circuit and whose columns correspond to discrete time segments of its execution; each entry is an integer identifying the physical network qubit to which that circuit qubit is assigned during that segment. In \emph{dynamic} partitionings, a qubit's assignment may change from segment to segment, whereas in \emph{static} partitionings it remains fixed, so each row contains a single repeated value.

The partitioner is deliberately modular: because every algorithm consumes the same two inputs and emits the same output format, a new partitioning algorithm can be integrated into the toolchain with minimal effort, streamlining the prototyping of new compilation strategies. This version of the software provides two primary partitioning algorithms, together with three additional benchmarking algorithms that establish a baseline for comparison and evaluation. We describe the two primary algorithms below; we emphasize that they are not proposed as novel methods, but rather as adaptations of established partitioning approaches to a hardware- and topology-aware setting.

\subsubsection{Dynamic Interaction Partitioning Algorithm}
\begin{figure}
    \centering
    \includegraphics[width=0.9\linewidth]{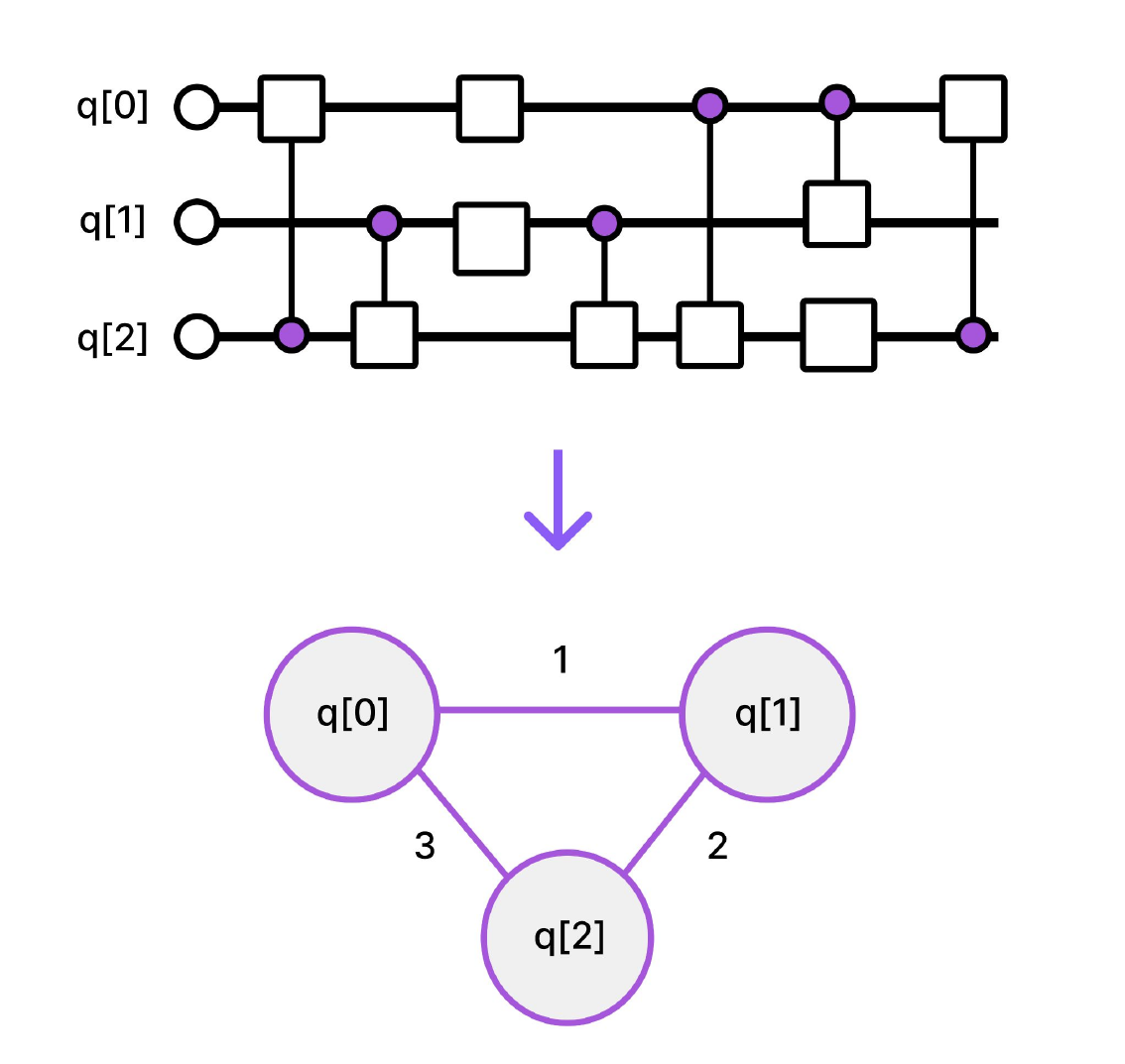}
    \caption{Interaction graph representation of a quantum circuit. Each node represents a qubit, and edge weights represent the number of two-qubit interactions between each pair of qubits.} \label{fig:interaction-graph}
\end{figure}

The Dynamic Interaction partitioning algorithm represents a quantum circuit as an
interaction graph, where vertices correspond to logical qubits and weighted
edges encode the number of two-qubit interactions between pairs of qubits
\cite{Zomorodi2018}. An example of this representation is shown in
Fig.~\ref{fig:interaction-graph}. This graph representation is useful because
it allows standard graph-partitioning algorithms to be applied to the circuit
mapping problem. In a graph-partitioning algorithm, the goal is to produce a series of `cuts' in the graph which divide the nodes into groups while minimizing the total weight of the cut edges. Partitioning the interaction graph of the full circuit produces a valid \emph{static} assignment of logical qubits to QPUs, in which each qubit remains on the same QPU throughout execution. Because each qubit stays fixed, no state teleportation is needed; however, a single static assignment cannot adapt to how the circuit's interaction pattern shifts over the course of execution, often forcing a large number of remote gates to bridge qubits that a different, time-varying placement could have kept local. To let the placement evolve over time, Kaur et al. \cite{cisco2025} proposed a dynamic partitioning algorithm that divides the circuit into time segments and partitions the interaction graph of each segment independently, introducing state teleportation operations to relocate qubit states between segments in addition to executing remote gates.

Our algorithm adapts this segment-based approach; its key distinction is that partitioning is performed in a \emph{network topology-aware} setting, enabling partitioning across networks of non-uniform QPUs with varying qubit counts and connectivity. The algorithm first selects a segment size based on the input circuit and divides the circuit into segments by two-qubit gate count. Only qubits participating in a two-qubit gate within a segment are included in that segment's interaction graph. The first segment is partitioned using an adaptation of the Kernighan-Lin (KL) heuristic \cite{kl1970}. The original KL algorithm produces a balanced \emph{bipartition} of a graph; we extend it in two ways to fit the distributed setting. First, rather than splitting the graph into two, we partition it into $k$ parts, one per QPU in the network. Second, we replace KL's balanced-size
objective with a capacity constraint: the network topology is parsed into a
capacity vector encoding the number of available qubits per QPU, and the
partitioner is constrained so that every part strictly respects its QPU's
capacity. Because QPUs may differ in size, these parts are in general unequal,
unlike the equal-sized halves KL targets by default. Each subsequent segment is then partitioned in turn, and a topology-aware cost function decides whether to adopt the new partition or retain the current one. This cost is measured in EPR pairs and accounts not only for the remote two-qubit gates required by the segment, but also for the cost of teleporting qubit states from their current QPU assignments to the new ones.

Because the framework admits arbitrary inter-QPU connectivity, the cost function does not assume all-to-all connectivity between QPUs. When two QPUs are not directly connected, the cost of relocating a qubit between them is computed over a route through intermediary QPUs, selected by a path-finding algorithm that minimizes total cost; where an intermediary QPU would exceed its capacity along that route, the algorithm inserts the state teleportation operations needed to keep it within capacity. Every inter- and intra-QPU operation required to realize a candidate partition is therefore counted explicitly. This explicit accounting allows the algorithm to evaluate candidate partitions under arbitrary network topologies: compilers that implicitly assume full connectivity cannot accurately estimate the cost of a partition on a constrained topology such as a chain, whereas our cost function estimates its operational cost under the specified network model. This topology-awareness is not limited to inter-QPU connectivity: the framework also accounts for the intra-QPU topology of each processor, which determines how data qubits must be routed to communication qubits when realizing remote operations during circuit reconstruction.

The segment length $\ell$ controls how often the qubit placement may change. A circuit containing $g$ two-qubit gates is divided into $\lceil g/\ell \rceil$ segments. A small $\ell$ permits frequent placement changes and may therefore increase the frequency of state teleportation, whereas $\ell \geq g$ produces a single static placement. Users may set $\ell$ directly to control the trade-off between placement flexibility and state-teleportation overhead. Otherwise, the default is
\begin{equation}
\ell = \operatorname{clip}\!\left(
    \left[2\sqrt{g}\right],
    \ell_{\min},
    \ell_{\max}
\right),
\label{eq:segment-length}
\end{equation}
where $[\cdot]$ denotes rounding to the nearest integer, $\operatorname{clip}(x,a,b)$ constrains $x$ to the interval $[a,b]$, $\ell_{\min}=10$ (or $1$ when $g<10$), and $\ell_{\max}=\min(100,g)$. This simple heuristic is intended to balance the number of segments against their size; it is not guaranteed to be optimal, and the best value of $\ell$ depends on both the circuit and the network topology, as discussed in Appendix~\ref{app:segment-sensitivity}. All results in this paper use the default defined in Eq.~\eqref{eq:segment-length}.

The two costs compared in Algorithm~\ref{alg:interaction-partitioning} are measured in EPR pairs. $\textsc{RemoteCost}$ sums, for each interaction-graph edge connecting qubits assigned to different QPUs, the pairs consumed by that edge's remote gates: one pair per gate between directly connected QPUs, and $1+2k$ pairs when the QPUs are separated by $k$ intermediates, as one operand is teleported along the route and returned. $\textsc{MoveCost}$ prices the relocations needed to reach the candidate placement; because $\textsc{Repartition}$ moves only active qubits and preserves each QPU's load, every relocation is a remote SWAP, consuming $2h$ pairs over $h$ hops---two per hop, one for each state teleportation. Each hop must provide two disjoint communication links, and a candidate with no valid route is assigned infinite cost to ensure it is not selected. The candidate is accepted if its remote-gate and relocation costs together are less than the remote-gate cost of retaining the current placement. The algorithm iterates over each segment, either accepting the new partition or retaining the previous one, until all segments have been processed. The final output is a partition schedule that specifies the QPU assignment of each circuit qubit at each segment. This result is then passed to the circuit reconstruction module, where it is resolved to an executable distributed quantum circuit. The full algorithm is described in pseudocode in Algorithm~\ref{alg:interaction-partitioning}.

\begin{algorithm}[t]
\caption{Dynamic Interaction partitioning}
\label{alg:interaction-partitioning}
\begin{algorithmic}[1]
\Require Circuit $C$, network $N$, segment length $\ell$
\Ensure Partition schedule $S = [P_1, \dots, P_m]$

\State Split $C$ into segments $C_1, \dots, C_m$, each containing at most $\ell$ two-qubit gates.
\State Let $\mathbf{b}$ denote the per-QPU qubit capacities of $N$.
\State Let $G_1$ be the interaction graph of $C_1$.
\State Set $P_1 = \textsc{Partition}(G_1, \mathbf{b})$.
\State Initialize $S = [P_1]$.

\For{$i = 2$ \textbf{to} $m$}
    \State Let $P_{\mathrm{old}} = P_{i-1}$.
    \If{$C_i$ contains no two-qubit gates}
        \State Set $P_i = P_{\mathrm{old}}$.
    \Else
        \State Let $G_i$ be the interaction graph of $C_i$.
        \State Set $P_{\mathrm{new}} = \textsc{Repartition}(G_i, P_{\mathrm{old}})$.
        \Comment{Only active qubits may move; QPU loads are preserved.}
        \State Set $c_{\mathrm{old}} = \textsc{RemoteCost}(G_i, P_{\mathrm{old}}, N)$.
        \State Set $c_{\mathrm{new}} =
        \textsc{RemoteCost}(G_i, P_{\mathrm{new}}, N)
        + \textsc{MoveCost}(P_{\mathrm{old}}, P_{\mathrm{new}}, N)$.
        \If{$c_{\mathrm{new}} < c_{\mathrm{old}}$}
            \State Set $P_i = P_{\mathrm{new}}$.
        \Else
            \State Set $P_i = P_{\mathrm{old}}$.
        \EndIf
    \EndIf
    \State Append $P_i$ to $S$.
\EndFor

\State \Return $S$
\end{algorithmic}
\end{algorithm}

\subsubsection{Hypergraph Partitioning Algorithm}
\label{sec:hypergraph}
Another common approach to partitioning distributed circuits is based on replacing the interaction graph representation with a hypergraph. A hypergraph is a graph in which an edge can connect more than two nodes, thus forming a \emph{hyperedge}. Each gate packet is rooted on a single \emph{control} qubit, which is shared with the remote QPU via one Cat-Entangler operation and released by a single Cat-Disentangler operation, so that every gate admitted to the packet is covered by the same entangled pair. Using these protocols, we construct distributed gate `packets' according to the following criteria \cite{wu2023, AndresMartinez2024, Burt2024}:

\begin{itemize}
    \item \textbf{Shared control.} A two-qubit gate may join a packet only if it shares a common control qubit with the packet's root.
    \item \textbf{Symmetric gates.} Because the controlled-$Z$ gate is symmetric in its control and target, a \texttt{cz} gate is admitted whenever the packet's root qubit appears as \emph{either} the control or target.
    \item \textbf{Control-qubit single-qubit gates.} A single-qubit gate applied to the root control qubit is permitted within a packet only if it is diagonal or anti-diagonal in the computational basis.
    \item \textbf{Target-qubit operations.} Single-qubit gates acting on the target qubits of a packet do not affect the shared control link and so do not break the packet.
    \item \textbf{Non-adjacent gates.} A gate that shares the root control does not need to come immediately after the previous gate in the packet. If another two-qubit gate appears in between, the later gate can still be added to the packet as long as that intervening gate acts on a separate set of qubits, since the two gates can then be reordered without changing the circuit. This lets a packet collect compatible gates that are spread out across the circuit, rather than only those that occur back-to-back.
\end{itemize}

\begin{figure}
    \centering
    \includegraphics[width=0.8\linewidth, trim=0 2cm 0 0]{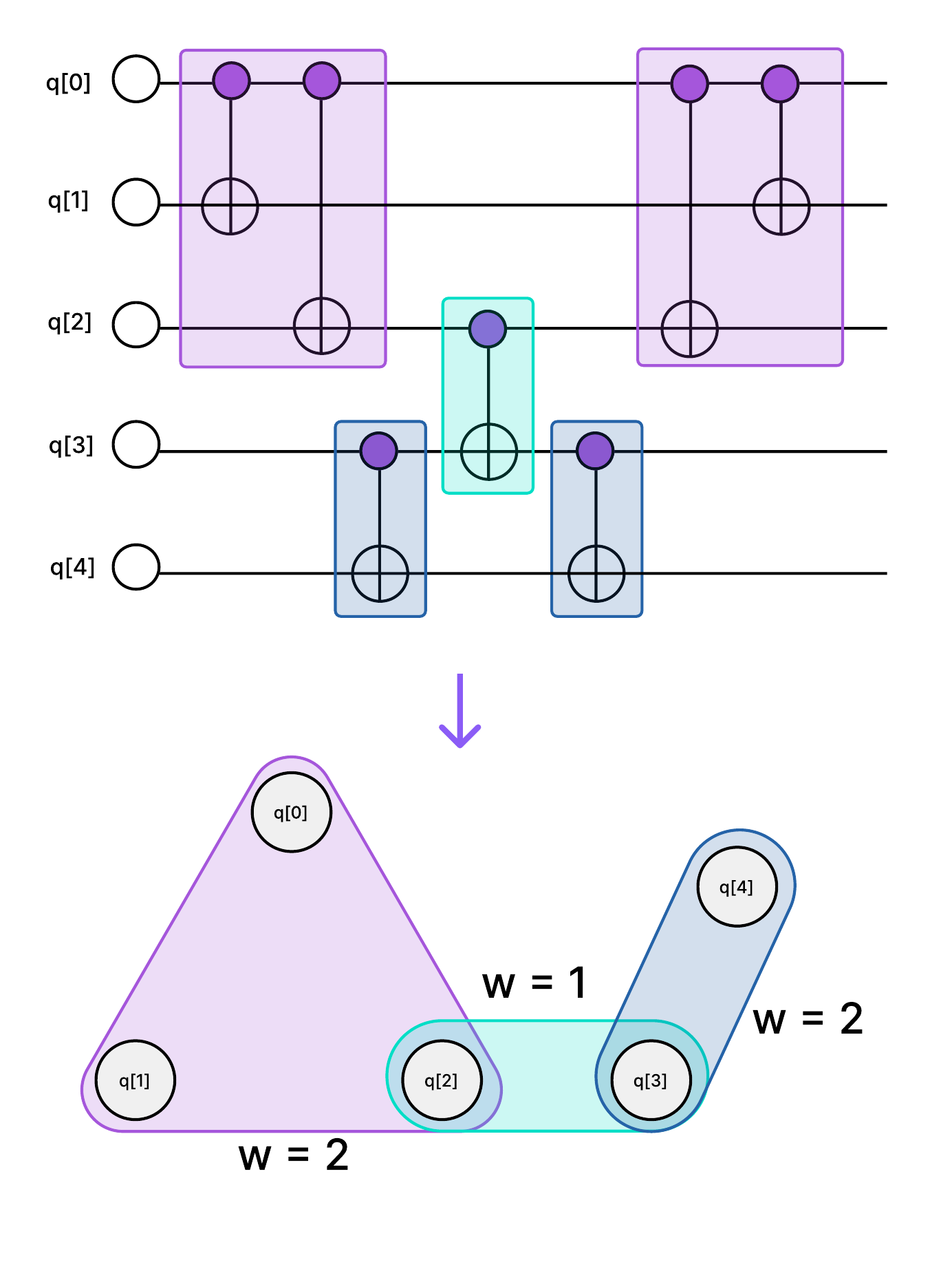}
    \caption{Hypergraph representation of a quantum circuit. Each node represents a qubit, and each hyperedge represents a gate packet connecting all qubits acted on by gates in that packet.} \label{fig:hypergraph}
\end{figure}

Because gates within a remote-gate group share an entangled pair, maximizing valid grouping opportunities during partitioning can reduce EPR-pair consumption. One approach to this is modeling the input circuit as a hypergraph, where each qubit is represented as a vertex, and each gate packet is represented as a hyperedge connecting all qubits acted on by gates in that packet, as shown in Fig.~\ref{fig:hypergraph}. Each hyperedge initially has a weight of 1, but if multiple gate packets act on the exact same set of qubits, the weight of the corresponding hyperedge is incremented. As a result, during partitioning, the cost of cutting a hyperedge is representative of the true number of EPR pairs that will be consumed by the created remote operations; without using hyperedges, the partitioner would overestimate this cost. It is worth noting that the groups formed while constructing the hypergraph will not necessarily be the same groups formed during circuit reconstruction.

After its construction, the hypergraph can be partitioned using any standard hypergraph partitioning algorithm. In this approach, we utilize the KaHyPar hypergraph partitioner \cite{KaHyPar2023}. As with the Dynamic Interaction partitioning algorithm, the Hypergraph partitioner is topology-aware; the number of partitions is equal to the number of QPUs in the network, and the KaHyPar partitioner is given partition-size constraints based on the qubit capacity for each specific QPU by adjusting the maximum \emph{block weight} parameter in the KaHyPar configuration. The output of the Hypergraph partitioner is a mapping of each qubit to a specific QPU, which is then used to construct a partition schedule for the entire circuit. It is important to note that the Hypergraph partitioner is not dynamic in time; therefore, state teleportation is not used as qubits are fixed to specific QPUs throughout the execution of the circuit. The full algorithm is described in pseudocode in Algorithm~\ref{alg:hypergraph-partitioning}.

\begin{algorithm}[t]
\caption{Hypergraph partitioning}
\label{alg:hypergraph-partitioning}
\begin{algorithmic}[1]
\Require Circuit $C$, network configuration $N$, optional segment length $\ell$
\Ensure Partition schedule $S$

\State Determine the QPU set $\mathcal{Q}$ from $N$.
\State Determine the QPU capacity vector $\mathbf{b}$ from $N$.
\State Initialize an empty packet multiset $\mathcal{P}$.

\For{each operation $g$ in $C$}
    \If{$g$ is a two-qubit gate and has not already been assigned to a packet}
        \State Initialize a packet $p = \{g\}$.
        \State Identify subsequent compatible gates that can be grouped with $g$.
        \State Add the grouped gates to $p$.
        \State Add $p$ to $\mathcal{P}$.
    \EndIf
\EndFor

\State Initialize an empty hypergraph $H = (V, E)$.
\State Set $V$ to the set of logical qubits appearing in $\mathcal{P}$.

\For{each packet $p \in \mathcal{P}$}
    \State Let $e_p$ be the set of logical qubits acted on by gates in $p$.
    \State Add hyperedge $e_p$ to $E$.
    \State If an identical hyperedge already exists, increment its weight.
\EndFor

\State Set $A = \textsc{HypergraphPartition}(H, \mathbf{b})$.
\State Assign any logical qubits not appearing in $H$ to QPUs with remaining capacity.

\If{$\ell$ is provided}
    \State Divide $C$ into operation segments $\{C_i\}_{i=1}^{m}$, each containing up to $\ell$ two-qubit gates.
\Else
    \State Let $\{C_i\}_{i=1}^{m} = \{C\}$.
\EndIf

\State Initialize $S = [\,]$.
\For{$i = 1$ to $m$}
    \State Append assignment $A$ to $S$.
\EndFor

\State \Return $S$
\end{algorithmic}
\end{algorithm}

\subsection{Circuit Reconstruction}
\label{subsec:reconstruction}
The \emph{Circuit Reconstruction} module receives the partition schedule produced by the partitioning module as its core input, alongside the network topology and original circuit. The purpose of this module is to convert these inputs into a valid distributed quantum circuit which is logically equivalent to the input circuit while being executable on the specified quantum network. The compiler converts this program into the graph representation used during compilation, producing a \emph{distributed circuit DAG}.

The first step of this process is to identify all necessary remote gate operations. This is done by stepping through the monolithic circuit and resolving each multi-qubit operation against the partition schedule. If a multi-qubit operation acts on qubits that are assigned to different QPUs, the DAG node for this operation is marked as a remote operation.

Next, the module resolves dynamic qubit placement via state teleportation. For dynamic partitioning schedules, the module steps through the partition schedule segment by segment. For each segment in which the QPU assignment of a circuit qubit changes, the module inserts a state teleportation operation to move the qubit from its current QPU to its new QPU. The current version of this software assumes qubit occupancy per QPU remains constant throughout the execution of the circuit; therefore, a state teleportation operation from QPU A to B is always accompanied by a state teleportation operation from QPU B to A to maintain the qubit count. The compiler makes no assumptions regarding the availability of buffer or memory qubits. We abstract this dual state teleportation as a \emph{remote swap} operation, which serves as a placeholder primitive, costing two EPR pairs---one per state teleportation. For qubit movements between QPUs that do not share direct connections, the module finds the shortest path through the network topology to route the qubit through intermediary QPUs, inserting remote swap operations along the path while ensuring that no QPU exceeds its qubit capacity at any point in time. These remote swap operations are inserted as new nodes in the appropriate location of the original DAG.

After resolving state teleportation operations, the module forms gate packets according to the gate-grouping criteria of the Hypergraph partitioner described previously. Stepping through the DAG, now annotated with remote operations and augmented with the inserted state teleportation operations, it identifies all gate packets that can be formed. Packet formation begins by inserting a Cat-Entangler operation immediately before the first remote gate encountered in the DAG, thereby initializing a packet; the module then extends the packet with subsequent gates that satisfy the grouping criteria. Upon encountering a gate that violates the criteria, and therefore terminates the packet, it inserts a Cat-Disentangler operation to close the packet. Both the Cat-Entangler and Cat-Disentangler are inserted as new nodes in the DAG, and every remote gate belonging to the packet is annotated with a unique packet identifier.

Executing a remote gate additionally requires that each involved data qubit be adjacent to a communication qubit on its own QPU, since the Cat-Entangler acts between a data qubit and a local communication qubit. To satisfy this, the module makes use of the intra-QPU connectivity specified in the network topology: if a data qubit involved in a remote gate is not directly connected to an available communication qubit, it is routed to one along the shortest path through that QPU's local connectivity graph, with a local \texttt{swap} gate inserted as a new DAG node for each hop along the path. Consequently, the compiled program respects not only the inter-QPU topology but also the internal connectivity of each individual QPU, and remote operations are only ever emitted between qubits that are physically able to interact. To this end, the module maintains a live mapping from each logical circuit qubit to the physical network qubit it currently occupies. As each operation is emitted, its operands are translated through this mapping so that the resulting gate acts on the correct physical qubits, and after each intra- or inter-QPU qubit movement the mapping is updated to reflect the new locations of the qubits before subsequent operations are processed. This ensures that, even as qubits move between and within QPUs over the course of execution, the operations emitted between physical qubits faithfully reflect the gate dependencies of the original logical circuit.

Finally, the distributed DAG is serialized to a standard OpenQASM format. Each QPU is assigned a unique qubit register, with each qubit in the register corresponding to a physical computation qubit on that QPU. We make the distinction between computation and communication qubits within each QPU; a unique per-QPU register for communication qubits is also created. The compiler inserts custom gates for each remote gate, `remote swap', Cat-Entangler, and Cat-Disentangler operation, which are output in a \texttt{.inc} file alongside the OpenQASM output. Serializing directly from the distributed DAG allows the compiler to remain modular in two respects. First, because the network operations are emitted as named custom-gate definitions in a separate include file rather than being expanded inline, the abstract distributed program is decoupled from the physical implementation of each primitive: an alternative realization of a remote gate, remote swap, or teleportation protocol can be substituted simply by redefining it in the \texttt{.inc} file, leaving the emitted program unchanged. Second, because the output conforms to the standard OpenQASM format, the distributed program interoperates directly with the surrounding toolchain---including the scheduler, circuit simulators, and verification tools---without requiring any bespoke intermediate representation. Adding a new remote primitive therefore requires only a corresponding DAG node type and gate definition. An example of a small input circuit alongside its compiled distributed form is shown in Fig.~\ref{fig:qasm-compile} in the Appendix.

The distributed compilation process inserts both local routing operations (\texttt{swap} sequences) and remote operations into the quantum circuit, necessitating verification that the compiled program preserves the original program's measurement-outcome distribution. The verification module reconstructs a monolithic circuit by replacing remote gates and remote swaps with their local equivalents and removing the Cat-Entangler and Cat-Disentangler scaffolding and communication registers. It then uses Qiskit circuit simulators to compare the original and reconstructed circuits in one of two modes: a deterministic mode that derives the complete output distributions from their pre-measurement statevectors and requires agreement within numerical tolerance, or a sampling-based mode that estimates the distributions over a finite number of shots and requires their Hellinger fidelity to exceed a user-specified threshold. To account for qubit permutation during compilation, the module aligns the measurement outcomes using the compiler's circuit-to-physical qubit mapping. This procedure verifies equality of measurement-outcome distributions rather than full unitary equivalence and treats the remote primitives themselves as correct by construction. Because statevector simulation requires storing $2^n$ amplitudes, verification is currently limited to circuits of up to 28 qubits; more scalable equivalence-checking techniques are left for future work.

\subsection{Scheduling}

After the circuit has been compiled into a distributed version of the program, our software provides a scheduler that assigns each distributed operation to a discrete time step. The scheduler takes as input the distributed circuit DAG, thus remaining language-agnostic. To build a schedule, users first select a pre-built hardware profile, or build their own. The hardware profile is currently distinct from the scheduler and parameterizes gate times and entanglement generation rates via a selected modality. The user can provide arbitrary one- and two-qubit gate times, measurement times, and entanglement generation rates for the desired hardware. Currently, this implies homogeneous modalities across a network that may be heterogeneous with regard to topology (i.e., varying QPU sizes and connectivities), though future work will include the ability to provide distinct hardware profiles for each node in the network.

The rates and times specified in the hardware profile are used to assign durations to each operation in the compiled circuit. Local single-qubit gates, two-qubit gates, and measurements are each assigned their corresponding time from the hardware profile. The entanglement primitives are assigned composite durations that reflect the local operations they comprise: a Cat-Entangler is timed as one two-qubit gate, one single-qubit gate, and one measurement, while a Cat-Disentangler is timed as two single-qubit gates and one measurement. A remote gate is therefore not a single fixed cost. A gate packet is scheduled as its Cat-Entangler, followed by one remote two-qubit gate for each gate grouped into the packet, followed by its Cat-Disentangler; the total duration of the packet thus grows with the number of gates it contains. The scheduler then uses these durations to build an execution schedule, which it outputs as a Gantt chart. The scheduler provides two modes. The deterministic mode uses the nominal mean duration $1/\lambda$. The discrete-event mode treats herald arrivals as a Poisson process of rate $\lambda$ and advances in intervals $\Delta t$, where $\Delta t$ is simply the simulation time interval, not a physical attempt duration. Each interval succeeds with probability $p=1-e^{-\lambda\Delta t}$; therefore, the number of intervals to success is geometric. The corresponding mean, $\Delta t/p$, approaches $1/\lambda$ for small $\Delta t$. Thus, the underlying process is Poissonian, while its interval-based simulation is geometric. Individual runs vary, and only one entanglement request can use a link at a time.

\begin{figure*}[!t]
    \centering
    \includegraphics[width=\textwidth]{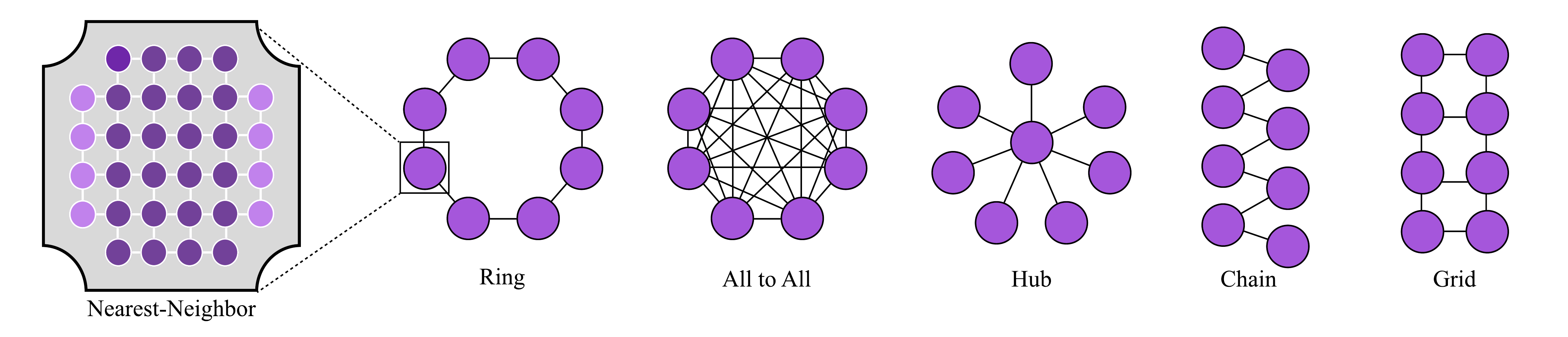}
    \caption{(\textit{Left}) Illustration of nearest-neighbor intra-QPU
    topology as used in evaluation. (\textit{Right}) Sample network
    topologies used for evaluation of the compiler framework.}
    \label{fig:topologies}
\end{figure*}

The deterministic scheduler is implemented as a simple greedy first-in-first-out (FIFO) algorithm that walks the distributed circuit DAG layer by layer in topological order. For each qubit, the scheduler tracks the earliest time at which it next becomes available; an operation is then assigned the earliest start time at which all of its qubits are simultaneously free, after which the availability time of each participating qubit is advanced by that operation's duration. Because operations are placed greedily in dependency order without any global optimization, the scheduler produces, in a single pass, a valid schedule that respects all circuit dependencies and per-qubit resource constraints. Entanglement generation time is constant and deterministic, given by the inverse of the generation rate.

The discrete-event schedulers instead model each remote operation as a request for an entangled pair on one of the network's communication links. A remote operation blocks until an entangled pair is successfully heralded on its link, at which point the Cat-Entangler and the remote gate may proceed. When several remote operations request the same link at once, the scheduler must decide which request to serve first, and it is precisely this link-arbitration policy that distinguishes the three discrete-event schedulers provided by the framework. All three share the same underlying simulation engine and differ only in how they break contention for links.

The first policy resolves multiple link requests in first-in-first-out order: when a link becomes available, the request that has been waiting longest is served. This treats all remote operations equally and serves as the discrete-event analogue of the deterministic FIFO scheduler. The second policy is a shortest-duration strategy: when a link becomes available, it serves the request whose remote operation has the shortest deterministic duration $d_o$, with ties broken by FIFO. Prioritizing short operations minimizes the average time requests wait for a link, at the cost of potentially delaying longer operations under heavy request traffic. The third strategy, referred to as the `critical-path algorithm,' prioritizes the request with the largest remaining weighted path length. Each operation $i$ is assigned a remaining-path cost $c_i = d_i + \max\{\, c_j : j \in \mathrm{succ}(i) \,\}$, where $d_i$ is its duration and $\mathrm{succ}(i)$ is the set of operations that directly depend on $i$ in the DAG; terminal operations, which have no successors, take $c_i = d_i$. When multiple requests contend for a link, the one with the largest $c_i$ is served first.



\section{Results}
\label{sec:results}

The configurability of our software stack facilitates the exploration of the wide design space of distributed quantum computing. Executing distributed quantum computing at scale will require co-design of hardware and software; researchers will need to explore the impact of network topology, intra-QPU connectivity and qubit count, while also considering compilation and scheduling strategies. Of course, these decisions must be made in the context of the quantum program to be executed. Our stack enables this exploration and demonstrates the effect of each of these design choices on the performance of a distributed quantum system. In this section, we present several demonstrations that can be performed using our software stack. We demonstrate that progress in the field will require a co-design approach, where hardware and software design choices are considered in tandem, rather than a `one-size-fits-all' approach. We obtain results using a variety of common quantum benchmark circuits via QASMBench~\cite{qasmbench2022}.

\subsection{Evaluation Methodology}

Throughout these demonstrations we make use of various network configurations with regard to two aspects: the \emph{inter-QPU} topology, which specifies how the QPUs are connected to one another, and the \emph{intra-QPU} topology, which specifies the connectivity of the computation qubits within each QPU. For the inter-QPU topology, throughout our results we utilized five representative QPU arrangements, illustrated in Fig.~\ref{fig:topologies}: a \emph{chain}, in which each QPU connects only to its immediate neighbors; a \emph{ring}, which closes the chain into a loop so that the end QPUs are also connected; a \emph{hub}, in which a single central QPU connects to all others; \emph{grid}, where QPUs are arranged in a nearest-neighbor fashion, and \emph{all-to-all}, in which all QPUs are connected to one another. For the intra-QPU topology, we consider both nearest-neighbor connectivity as shown in Fig.~\ref{fig:topologies} and full all-to-all connectivity, in which all qubits are connected to each other, including communication qubits.

To contextualize the performance of our two primary partitioners---the Dynamic Interaction partitioner and the Hypergraph partitioner---our software provides three additional partitioning algorithms for benchmarking purposes. The \emph{Static Interaction} partitioner applies the same Kernighan--Lin interaction-graph partitioning as the Dynamic Interaction partitioner, but computes a single assignment for the entire circuit rather than re-partitioning over time; this produces a distributed program without state teleportation. The \emph{static benchmark} partitioner assigns logical qubits to QPUs sequentially, filling each QPU to capacity without any analysis of the circuit's interaction structure, and serves as a naive, low-effort baseline. For example, for a 10-qubit circuit to be run on two 5-qubit QPUs, the first five qubit indices in the circuit would be assigned to the first QPU, and the remaining five to the second. The \emph{random benchmark} partitioner fills the QPUs to the same capacities, but chooses \emph{which} qubits land on each QPU by shuffling the logical qubit indices uniformly at random rather than assigning them in order. All three baselines produce a single static assignment that respects the per-QPU qubit capacities. For every benchmark of $28$ qubits or fewer, we used the verification module described in the \emph{\nameref{subsec:reconstruction}} section to confirm that the compiled distributed program reproduces the input circuit's output distribution, for each partitioning strategy and network configuration reported below.
\subsection{Dependence on Inter-QPU Topology}
\begin{figure}
    \centering
    \includegraphics[width=\linewidth, trim= 0 0 0 10mm, clip]{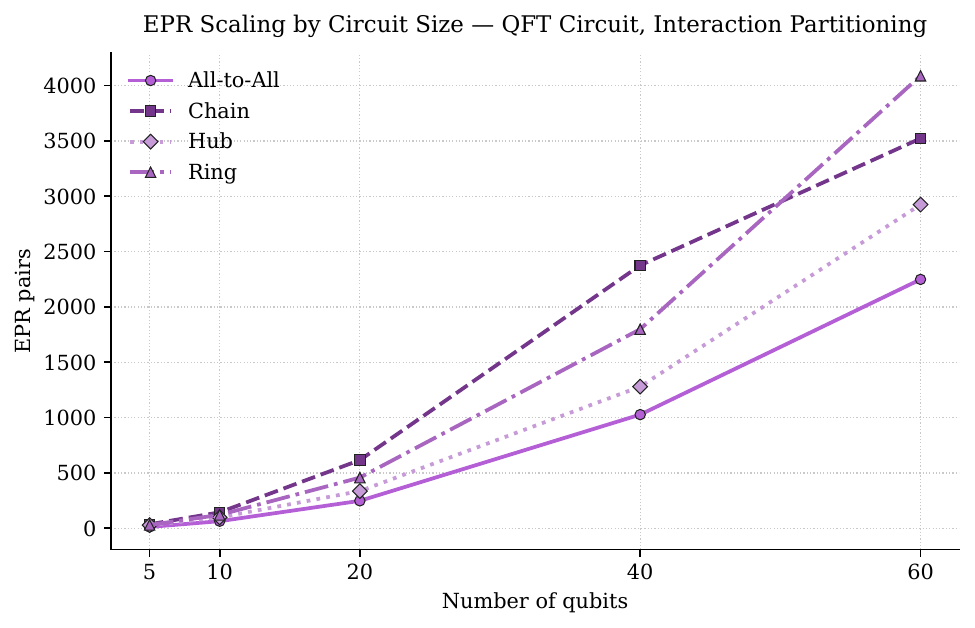}
    \caption{\textbf{EPR-pair cost by circuit size, across inter-QPU topologies.}} EPR-pair consumption as a function of circuit size for Quantum Fourier Transform (QFT) circuits compiled with the Dynamic Interaction partitioner and mapped onto a 5-QPU network under \emph{chain}, \emph{hub}, \emph{ring}, and all-to-all inter-QPU topologies, each with nearest-neighbor intra-QPU connectivity. The choice of network topology significantly affects EPR-pair cost, and the relative ordering of the topologies can change with circuit size. \label{fig:epr_scaling}
\end{figure}

\raggedbottom
In Fig.~\ref{fig:epr_scaling}, we explore the execution of Quantum Fourier Transform (QFT) circuits of increasing sizes across varying 5-QPU network topologies. Here, we consider nearest-neighbor connectivity within each QPU, with processors arranged in \emph{chain}, \emph{hub}, \emph{ring}, and \emph{all-to-all} topologies, shown in Fig.~\ref{fig:topologies}. These particular topologies demonstrate the impact of network configuration on the performance, as measured in EPR pairs consumed, of a common quantum algorithm. While all topologies examined here scale with the number of qubits in the circuit, we note that certain topologies perform significantly better than others depending on the qubit count. All-to-all, which is an idealized topology, performs best across all qubit counts. Notably, among the evaluated topologies, the chain has the highest cost at and below 40 qubits, but at 60 qubits it outperforms the ring. This crossover reflects a trade-off between routing distance and gate grouping. A remote operation between directly linked QPUs costs a single EPR pair, whereas one spanning an intermediary must be routed at higher cost; because the QFT interaction graph is effectively all-to-all, cost is governed by the inter-QPU routing distances. At smaller sizes the ring keeps every QPU within two hops of every other, versus up to four hops on the chain, making it the cheaper topology. As the circuit grows, however, the ring's inexpensive qubit movement leads the dynamic partitioner to favor frequent state teleportation, re-partitioning in most segments. This repeated reassignment fragments cat-entanglement groups and pushes remote operations onto distance-two links. On the chain, long-range teleportation is too costly to be worthwhile, so the placement freezes into a near-static layout that keeps gates grouped on adjacent QPUs. At 60 qubits the ring's group fragmentation outweighs its distance advantage, and its EPR cost overtakes the chain's (the ring forms 2756 remote-gate groups with 48\% requiring routing and consumes 4089 EPR pairs, versus 2496 groups, 41\% routed, and 3523 EPR pairs for the chain). While this is just a particular example, it illustrates the complex interplay between circuit structure and network topology, and the need to adapt compilation strategy to both.

Across most circuit sizes, increased inter-QPU connectivity is associated with lower EPR-pair consumption because it reduces the routing distance of remote operations and qubit movement. However, connectivity degree alone does not determine the cost: inter-QPU distances, centrality, circuit structure, and compiler placement decisions also contribute. The hub topology illustrates this: despite having no more inter-QPU links than the chain, it attains the second-lowest cost at every circuit size examined here, because its central QPU keeps every pair of processors within two hops. Across the evaluated QFT instances, all-to-all connectivity yields the lowest cost, while the relative performance of the sparse topologies depends on how the circuit is mapped onto the network.

\subsection{Dependence on Intra-QPU Connectivity}

\begin{figure}
    \centering
    \makebox[\linewidth][c]{\includegraphics[width=1.08\linewidth]{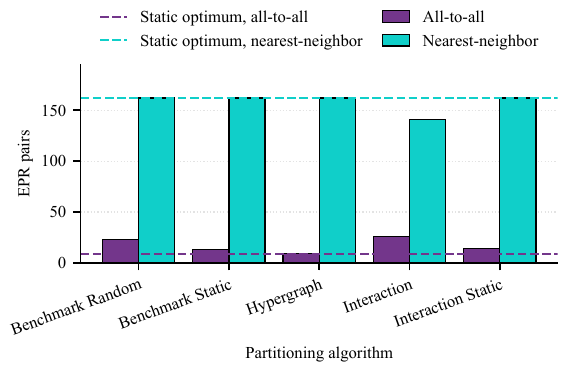}}
    \vspace{2mm}

    \makebox[\linewidth][c]{\includegraphics[width=1.08\linewidth]{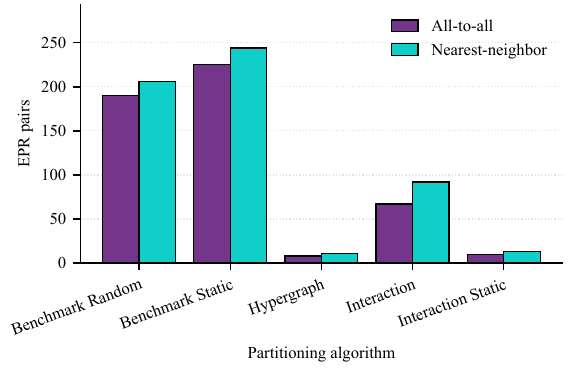}}
    \caption{\textbf{Cost of intra-QPU connectivity, for an 18-qubit QFT circuit (top) and a 64-qubit Adder circuit (bottom).} The cost, in EPR pairs, of changing the intra-QPU connectivity assumption from all-to-all to nearest-neighbor, for each circuit executed on a 2-QPU network. For each partitioning strategy, the paired bars give the EPR cost under all-to-all (left) and nearest-neighbor (right) intra-QPU connectivity. The dashed lines on the QFT panel mark the static optimum for each network, obtained by compiling every one of the $\binom{18}{9}$ balanced qubit assignments and taking the lowest cost; static here means the assignment is fixed for the whole circuit, so no qubit is moved between QPUs. The dynamic interaction partitioner is not bound by these lines, and on the nearest-neighbor network it falls below the static optimum by leveraging state teleportation.} \label{fig:intra_qpu_connectivity}
\end{figure}

In Fig.~\ref{fig:intra_qpu_connectivity}, we explore the impact of intra-QPU connectivity on the performance of distributed quantum computing. Here, we consider two common quantum algorithms, a 64-qubit Adder circuit and an 18-qubit QFT circuit, executed across a simple 2-QPU network. When the input network has nearest-neighbor intra-QPU connectivity, emulating processors such as superconducting systems, the EPR pair requirements are significantly higher than when the QPUs are assumed to have all-to-all connectivity, such as in trapped-ion systems. This effect is particularly significant in the QFT circuit, in which EPR cost is increased more than 10-fold in certain cases. In the Adder circuit, this effect is less pronounced, though every case incurs an additional EPR pair overhead when the assumption changes from all-to-all to nearest-neighbor connectivity. This result is due to the gate grouping protocol; when QPUs have sparse connectivity, the compiler must insert local SWAP operations to move qubits into positions adjacent to communication qubits to perform remote operations. Remote operations that require these SWAPs cannot share a cat-entanglement group, splitting what would otherwise be a single group into several. With all-to-all connectivity, no such SWAP operations are required. This permits larger gate groups and reduces EPR-pair requirements. The penalty is determined primarily by the extent to which SWAP insertion disrupts remote-gate grouping, rather than by the number of SWAPs alone. QFT is especially sensitive because its dense interaction pattern creates many remote gates that group well under all-to-all connectivity; nearest-neighbor routing breaks those groups, pushing the cost toward one EPR pair per remote gate and causing the partitioning strategies to converge. For QFT~(18), every balanced static assignment cuts $9\times9=81$ pairs in the circuit's complete interaction graph. The benchmark implements each such interaction with two CNOTs, so the loss of gate grouping produces $2\times81=162$ EPR pairs regardless of which balanced static assignment is chosen; the Dynamic Interaction partitioner falls below this value by moving qubits during execution. The Adder is less affected because its sparse, repetitive interaction structure admits little grouping to begin with, so nearest-neighbor routing has correspondingly less grouping advantage to erode.

The significant impact of intra-QPU connectivity on EPR pair consumption demonstrates the importance of allowing for full configurability of the software stack. Prior work often assumes all-to-all connectivity, which can yield overly optimistic results that do not generalize to all hardware architectures. By tailoring compilation to a given hardware configuration, users can make informed decisions about the design of distributed quantum systems based on the processors available to them. The top panel of Fig.~\ref{fig:intra_qpu_connectivity} specifically demonstrates this fact; while holding the inter-QPU topology and quantum circuit constant, the EPR pair requirements increased substantially when we changed from all-to-all intra-QPU connectivity to nearest-neighbor connectivity. For example, the Hypergraph partitioner's EPR-pair consumption increased by more than $10\times$ under this change.

\begin{figure*}
    \centering
    \includegraphics[width=0.85\textwidth]{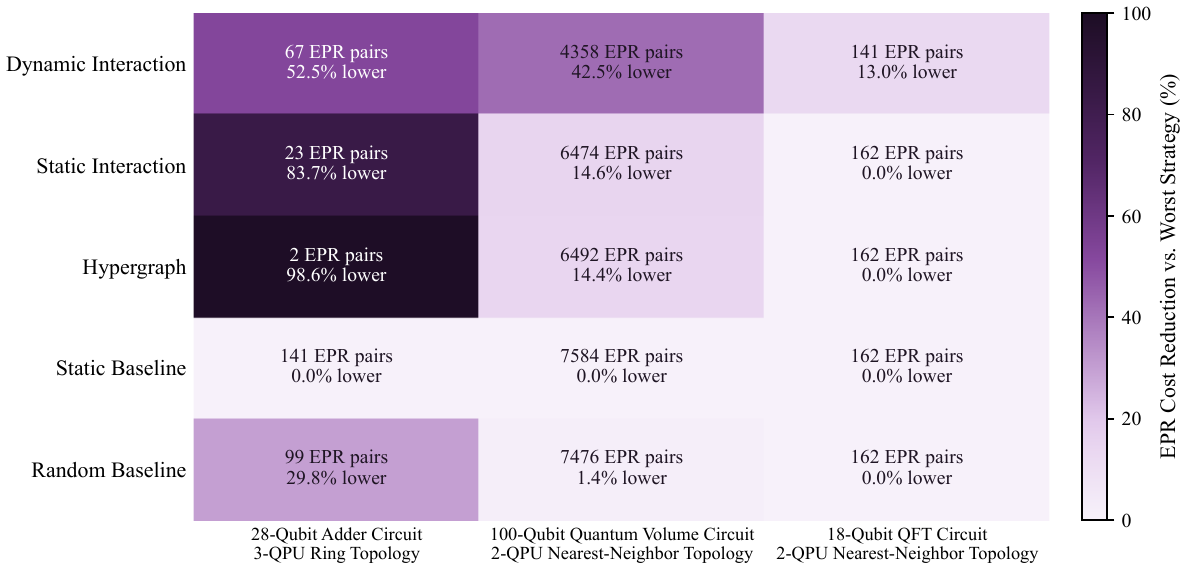}
    \caption{\textbf{Relative partitioning strategy performance.} A comparison of partitioning strategies' performance (in terms of EPR-pair consumption) across varying circuit and topology configurations. Each cell reports the EPR-pair count for one partitioning strategy, together with its percentage reduction relative to the worst-performing strategy in the same column. Color encodes this per-column relative reduction on a shared $0$--$100\%$ scale, with darker cells indicating larger reductions. Absolute counts for all circuit and network combinations are given in Tables~\ref{tab:epr-2qpu-all-to-all}--\ref{tab:epr-grid-4qpu}.} \label{fig:heatmap}
\end{figure*}

\subsection{Comparison of Partitioning Strategies}

{In Fig.~\ref{fig:heatmap}, we examine the performance of the various compiler algorithms across a sample of benchmark circuits and topologies. The results demonstrate that the best-performing compiler algorithm varies across the evaluated circuits and topologies. For example, the repetitive controlled-gate structure of the Adder circuit is well exploited by the Hypergraph partitioner, which excels at partitioning circuits to maximize gate grouping; on a 3-QPU ring topology it attains minimal EPR-pair consumption, whereas the Dynamic Interaction partitioner requires more than thirty times as many EPR pairs to execute the same circuit. For a Quantum Volume circuit executed on a basic 2-QPU network, however, the Dynamic Interaction} algorithm outperforms the Hypergraph algorithm by a large margin. On the 4-QPU grid---the only configuration in Tables~\ref{tab:epr-2qpu-all-to-all}--\ref{tab:epr-grid-4qpu} containing QPU pairs without a direct link---the Dynamic Interaction partitioner can lose this advantage because segment-to-segment reassignments may require multi-hop state teleportation, whose cost can outweigh the resulting remote-gate savings; Appendix~\ref{app:segment-sensitivity} shows that whether this overhead dominates depends strongly on the segment length. Although both optimized algorithms outperform the static and random baselines in these two cases, executing a QFT circuit on a basic 2-QPU network with nearest-neighbor intra-QPU connectivity yields identical performance across nearly all algorithms. This not only indicates the need for more advanced compiler strategies, but, more importantly, demonstrates the limitations of relying on a single compiler strategy across all instances. To effectively navigate the wide parameter space of distributed quantum computing, the algorithmic layer must adapt its strategy to the circuit and the network topology. Our software stack enables systematic exploration of this parameter space, yielding insights that can inform the design of optimized DQC systems.

\subsection{Dependence on Link-Arbitration Strategy}

\begin{figure}[t!]
    \centering
    \includegraphics[width=\linewidth]{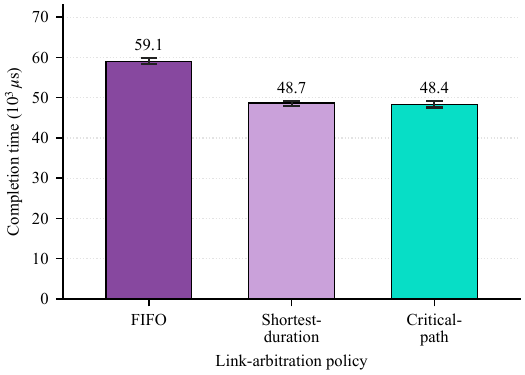}
    \caption{\textbf{Completion time by link-arbitration policy.} Mean completion time for three link-arbitration policies in the controlled fan-out contention experiment over 16 matched stochastic seeds. Error bars show $\pm1$ standard error of the mean after accounting for shared seed-to-seed variation. The experiment demonstrates sensitivity to scheduling strategy and is not a hardware-performance projection.} \label{fig:scheduler-divergence}
\end{figure}

To determine whether link arbitration can materially affect execution time, we conducted a controlled scheduler experiment rather than a hardware benchmark. The experiment uses a 3-QPU, 25-qubit fan-out circuit, shown in Appendix Fig.~\ref{fig:fanout-contention-circuit}. The circuit was constructed to create both link contention and dependent downstream work. Accordingly, FIFO produces the longest schedules, while both contention-aware policies shorten them by a comparable margin (Fig.~\ref{fig:scheduler-divergence}).

The separation requires both simultaneous link contention and remote operation segments of different durations: without queued entanglement requests, all policies make the same choice, while equal-duration segments cause shortest-duration to fall back to FIFO order. The schedulers otherwise share the same simulation engine and differ only in this arbitration rule. The hardware profile assigns durations of 63~$\mu$s to a single-qubit gate, 650~$\mu$s to a two-qubit gate, and 250~$\mu$s to a measurement, taken from published calibration data for the IonQ Forte Enterprise~1 trapped-ion processor~\cite{ionq_backends}, together with an EPR-generation rate of $2.5\times10^{-4}~\mu\mathrm{s}^{-1}$ ($250~\mathrm{s}^{-1}$)~\cite{PhysRevLett.133.090802}. These values correspond to a representative trapped-ion system, though we emphasize that all such parameters are configurable in the software, as this framework is modality-agnostic by design. The circuit and timing profile were intentionally selected to demonstrate that a small change in scheduling strategy can affect the result. Each policy was evaluated on the same 16 stochastic seeds. Because entanglement generation is stochastic, absolute completion times vary substantially from seed to seed, and the policies overlap heavily when compared in absolute terms. Comparing policies seed by seed removes this shared noise: relative to FIFO, shortest-duration reduces completion time by $17.7\pm1.3\%$ and critical-path by $17.9\pm1.5\%$ (mean $\pm$ s.e.m., $n=16$), and both policies outperform FIFO on every seed. The two contention-aware policies are indistinguishable from one another on this circuit ($0.1\pm1.6\%$), indicating that the benefit here comes from accounting for link contention at all rather than from the particular priority rule used.

\section{Discussion \& Conclusion}
\label{sec:discussion}

In this work, we presented an open-source, modular framework for the compilation and scheduling of distributed quantum programs. The framework spans the full path from program specification to executable schedule: the Quantum Network Constructor captures arbitrary inter- and intra-QPU topologies in a standard configuration format, the compiler partitions and reconstructs circuits into distributed programs that strictly respect the specified network, and the scheduler assigns operations to discrete time steps under either deterministic or stochastic entanglement generation models. By accepting and emitting standard OpenQASM, the framework extends, rather than replaces, the existing ecosystem of monolithic compilation and simulation tools, allowing distributed execution to be integrated within existing quantum compiler workflows.

A central conclusion of our results is that compilation-strategy performance depends on both the circuit and network configuration. The Hypergraph partitioner exploits the repetitive controlled-gate structure of certain circuits to achieve minimal EPR pair consumption through gate grouping, yet the Dynamic Interaction partitioner outperforms it by a large margin on comparatively unstructured circuits such as Quantum Volume in the 2- and 3-QPU cases; in other instances, the optimized algorithms offer little advantage over naive baselines. This observation argues strongly against a `one-size-fits-all' approach to distributed compilation. Rather than proposing a single dominant algorithm, we contend that the algorithmic layer of a distributed quantum software stack must adapt its strategy to the structure of the circuit and the topology of the network on which it is executed. The standardized partitioner interface provided by our framework is designed to support this adaptivity, allowing candidate strategies to be evaluated on a per-workload basis.

Our results further demonstrate that hardware assumptions embedded in the compilation process materially change its conclusions. While holding the circuit and inter-QPU topology fixed, changing the intra-QPU assumption of all-to-all connectivity to nearest-neighbor connectivity increased EPR pair consumption by more than an order of magnitude in the case of the 18-qubit QFT circuit (Fig.~\ref{fig:intra_qpu_connectivity}). Meanwhile, changing the choice of inter-QPU topology alone produced significant differences in resource requirements for identical workloads. Prior work that adopts idealized connectivity assumptions therefore risks producing resource estimates that are overly optimistic and that do not generalize across hardware architectures. Because our framework makes no inherent topology assumptions, it enables compilation results that reflect the operational cost on a given platform, and in doing so supports the hardware-software co-design methodology that we argue is necessary for distributed quantum computing. Decisions regarding interconnect topology, qubit connectivity, and compilation strategy are tightly linked, and must be evaluated in tandem rather than in isolation.

Although EPR-pair consumption is our primary compilation metric, it does not by itself determine execution time or the size and depth of circuits that a system can support. Lower EPR-pair requirements generally improve execution feasibility, but hardware viability also depends on the link fidelity, available communication and memory qubits, parallel-link capacity, qubit coherence times, and required output fidelity. The reported EPR counts should therefore be interpreted as measures of entanglement demand rather than hardware-feasibility limits. Future work will evaluate EPR consumption, runtime, and output fidelity jointly to determine which workloads satisfy specified time and fidelity requirements.

By standardizing the interfaces between these layers spanning quantum networking, compilation, and scheduling, the framework allows users to contribute within their domain of expertise---a new partitioning heuristic, an alternative remote-gate primitive, a link-arbitration policy---without reimplementing the surrounding stack. The included baseline algorithms additionally provide a consistent point of comparison, supporting reproducible evaluation of new compilation and scheduling strategies.

As distributed quantum hardware matures, the questions this framework is designed to answer---how many entangled pairs a workload requires, how that requirement scales with circuit size, topology, and connectivity, and which compilation strategy best suits a given platform---will directly inform the design of early distributed systems. We offer this framework as a foundation for that effort: a configurable, hardware-aware baseline upon which the community can build, benchmark, and co-design the software and hardware of distributed quantum processors.

\section{Future Work}
\label{sec:futurework}

Our software framework is intended to shorten the distance between high-level distributed quantum algorithms and execution on actual quantum processors with entanglement-based interconnects. Future work will further narrow this gap. Specifically, expansion of the framework to operate on a logical level by integrating distributed quantum error correction (QEC) will be a key step in this direction. As the quantum industry shifts from noisy intermediate-scale quantum (NISQ) devices to fault-tolerant quantum computing (FTQC), the ability to distribute quantum programs across the next generation of machines is critical.

In addition to extending the framework to support logical-level DQC, future work will focus on further optimizing the compilation and scheduling strategies proposed here. This initial work provides the toolkit needed to explore and implement enhanced algorithms that minimize the EPR-pair consumption of a distributed program. While improved partitioning strategies will be a key component, our results suggest that strategy performance depends strongly on the circuit and network topology, motivating support for multiple partitioning algorithms. Implementing a framework that can adaptively select the predicted best-performing strategy for a given circuit and network configuration—either through machine-learning approaches or more traditional heuristic methods—will be a key step in this direction. In addition, we will explore optimized scheduling strategies that make decisions beyond link arbitration. Such strategies may benefit from machine-learning approaches, including reinforcement learning, to optimize policies for stochastic entanglement generation.

Finally, the framework will be extended to lower layers of the stack to facilitate execution on hardware. The current DQC ecosystem lacks the standardized software-hardware interface available in monolithic quantum computing, through which users can run programs on quantum processors via the cloud. Building this interface and the standards necessary to translate high-level distributed quantum programs, specifically the entanglement-based network operations, into low-level hardware instructions is essential to make DQC a reality. This hardware-software interface will be supported through the development and integration of distributed quantum circuit simulators, which will allow for testing and validation of distributed programs on noisy processors and links.

\section*{Code Availability}
The framework described in this work is open source and freely available at \url{https://github.com/memQGit/dqc}. The software project is archived on Zenodo under the concept DOI \url{https://doi.org/10.5281/zenodo.22260966}; the version used in this work is DQC v0.1.2~\cite{dqc2026}.

\section*{Acknowledgment}
We are grateful to the memQ team for their engagement and support throughout this work.

\bibliographystyle{IEEEtran}
\bibliography{Bibliography}

\onecolumn
\appendices
\section{Software Framework Comparison}

Table~\ref{tab:software-comparison} compares the capabilities of representative distributed-quantum software frameworks. Open source is marked \emph{Yes} when publicly accessible source carrying an explicit open-source license was identified; \emph{No} means no such release was found as of August~31, 2026. We consider a \emph{time-resolved operation schedule} to be an output that assigns explicit start and end times to every local and network operation in a compiled distributed circuit. For libraries that are not open source, the capabilities reported here are inferred from their corresponding publications.

\begin{table}[H]
    \centering
    \caption{\upshape Capability comparison across representative distributed-quantum software frameworks.}
    \label{tab:software-comparison}
    \footnotesize
    \setlength{\tabcolsep}{3pt}
    \renewcommand{\arraystretch}{1.4}
    \begin{tabularx}{\textwidth}{@{}>{\raggedright\arraybackslash}p{2.0cm}>{\raggedright\arraybackslash}p{2.0cm}>{\raggedright\arraybackslash}p{1.55cm}>{\raggedright\arraybackslash}p{1.7cm}>{\raggedright\arraybackslash}p{1.7cm}>{\raggedright\arraybackslash}X>{\raggedright\arraybackslash}p{1.7cm}>{\raggedright\arraybackslash}p{2.2cm}>{\centering\arraybackslash}p{0.9cm}@{}}
        \toprule
        \textbf{Framework} & \textbf{Primary layer} & \textbf{Automatic circuit distribution} & \textbf{Inter-QPU/ inter-node connectivity} & \textbf{Local connectivity within each QPU} & \textbf{Other topology constraints} & \textbf{Produces a time-resolved operation schedule} & \textbf{Noise/decoherence model} & \textbf{Open source} \\
        \midrule
        SeQUeNCe \cite{Sequence2021} & Quantum-network discrete-event simulation & No & Configurable & Not modeled & Channel and node properties configurable & No & Yes & Yes \\
        NetSquid \cite{NetSquid2021} & Quantum-network and modular-system simulation & No & Configurable & Configurable & Component and channel properties configurable & No & Yes & No \\
        QuISP \cite{QuISP2022} & Quantum-repeater and network simulation & No & Configurable & Not modeled & Repeater and link properties configurable & No & Yes & Yes \\
        Qoala \cite{qoala2025} & Node execution environment and simulator & No & Configurable & Configurable & Node hardware and network resources configurable & No & Yes & Yes \\
        Pytket-DQC \cite{AndresMartinez2024} & Distributed-circuit compiler & Yes & Configurable & Not modeled & QPU capacity and communication-qubit capacity configurable & No & No & Yes \\
        Ferrari et al. \cite{ferrari2023} & Distributed-circuit compiler & Yes & Configurable & Configurable & Channel capacity and communication-qubit placement configurable & No & No & No \\
        Mengoni et al. \cite{mengoni2026} & Distributed-circuit compiler & Yes & Configurable & Not modeled & None & No & No & No \\
        Kaur et al. \cite{cisco2025} & Distributed-circuit compiler & Yes & Configurable & Not modeled & A cost can be assigned to each inter-QPU link & No & No & No \\
        This work (memQ DQC) & Distributed compiler and scheduler & Yes & Configurable & Configurable & Communication-qubit and routing constraints configurable & Yes & No & Yes \\
        \bottomrule
    \end{tabularx}
\end{table}

\clearpage
\section{Example Compiled Program}

Figure~\ref{fig:qasm-compile} shows a concrete example of the compilation workflow described in the Circuit Reconstruction section: a small monolithic input circuit (a) alongside the distributed program (b) produced by the compiler for a two-QPU network. The example illustrates the per-QPU computation and communication registers introduced by the compiler, and the realization of each inter-QPU gate as a remote CNOT wrapped in a Cat-Entangler/Cat-Disentangler pair. It is worth noting that this particular program does not contain any state teleportation operations.

\begin{figure}[H]
  \centering
  \begin{minipage}[t]{0.46\linewidth}
\begin{lstlisting}[style=qasm]
OPENQASM 3.0;
include "stdgates.inc";
qubit[6] q;
bit[6] b;
h q[0];
cx q[0], q[1];
h q[3];
cx q[1], q[2];
cx q[3], q[4];
t q[5];
cx q[4], q[5];
cx q[0], q[3];
h q[2];
cx q[2], q[5];
cx q[1], q[4];
b[0] = measure q[0];
b[1] = measure q[1];
b[2] = measure q[2];
b[3] = measure q[3];
b[4] = measure q[4];
b[5] = measure q[5];
\end{lstlisting}
    \centering{\small (a) Input circuit (monolithic)}
  \end{minipage}\hfill
  \begin{minipage}[t]{0.50\linewidth}
\begin{lstlisting}[style=qasm]
OPENQASM 3.0;
include "distgates.inc";
include "stdgates.inc";
qubit[3] q0;   // QPU 0 computation qubits
qubit[3] q1;   // QPU 1 computation qubits
qubit[1] c0;   // QPU 0 communication qubit
qubit[1] c1;   // QPU 1 communication qubit
bit[6] b;
h q0[0];
cx q0[0], q0[1];
h q1[0];
cx q0[1], q0[2];
cx q1[0], q1[1];
t q1[2];
cx q1[1], q1[2];
catent    q0[0], q1[0], c0[0], c1[0];
rcx       q0[0], q1[0], c0[0], c1[0];
catdisent q0[0], q1[0], c0[0], c1[0];
h q0[2];
catent    q0[2], q1[2], c0[0], c1[0];
rcx       q0[2], q1[2], c0[0], c1[0];
catdisent q0[2], q1[2], c0[0], c1[0];
catent    q0[1], q1[1], c0[0], c1[0];
rcx       q0[1], q1[1], c0[0], c1[0];
catdisent q0[1], q1[1], c0[0], c1[0];
b[0] = measure q0[0];
b[1] = measure q0[1];
b[2] = measure q0[2];
b[3] = measure q1[0];
b[4] = measure q1[1];
b[5] = measure q1[2];
\end{lstlisting}
    \centering{\small (b) Compiled distributed circuit}
  \end{minipage}
  \caption{Compilation of a toy six-qubit circuit. (a)~The input is a small,
  sample circuit of {four single-qubit gates (\texttt{h},
  \texttt{t}) interleaved with seven \texttt{cx} gates},
  written for a single six-qubit register. (b)~The circuit compiled for a
  two-QPU network in which each QPU holds three computation qubits and one
  communication qubit, with the two QPUs directly connected. The compiler
  replaces the single register \texttt{q} with per-QPU computation registers
  (\texttt{q0}, \texttt{q1}) and communication registers (\texttt{c0},
  \texttt{c1}). {CNOTs between co-located qubits remain local,
  while each \texttt{cx} whose operands are assigned to different QPUs is
  realized as a remote CNOT (\texttt{rcx}) wrapped in its own
  \texttt{catent}/\texttt{catdisent} pair that shares a single entangled pair;
  here, three of the seven \texttt{cx} gates become remote.}
  All custom gate definitions are emitted in the accompanying
  \texttt{distgates.inc} file.}
  \label{fig:qasm-compile}
\end{figure}

\clearpage
\section{EPR-Pair Distribution Cost}

Tables~\ref{tab:epr-2qpu-all-to-all}--\ref{tab:epr-grid-4qpu} report the full EPR-pair distribution cost achieved by each partitioning algorithm across a range of benchmark circuits and network topologies. Fig.~\ref{fig:topologies} illustrates the inter- and intra-QPU topology configurations represented in these tables. For each circuit, the ``Best Benchmark'' column reports the lower of the static-benchmark and random-benchmark costs. Every connected pair of QPUs is joined by exactly two inter-QPU links, so each QPU carries two communication qubits per neighbor. The two QPUs of a 2-QPU network are directly connected, so these networks differ only in their intra-QPU connectivity: Table~\ref{tab:epr-2qpu-all-to-all} uses all-to-all and Table~\ref{tab:epr-2qpu-nearest-neighbor} nearest-neighbor. In the 3-QPU case the QPUs are connected in a closed loop (`ring'), and in the 4-QPU case in a 2-by-2 grid; both use nearest-neighbor intra-QPU connectivity.


\begin{table}[H]
\centering
\caption{Distribution cost (EPR pairs consumed) on a 2-QPU network with all-to-all intra-QPU connectivity.}
\label{tab:epr-2qpu-all-to-all}
\begin{tabular}{lcccc}
\toprule
\textbf{Circuit (qubits)} & \textbf{Hypergraph} & \textbf{Dynamic Interaction} & \textbf{Static Interaction} & \textbf{Best Benchmark} \\
\midrule
Adder (28)       &        7 &       23 &        7 & 93 \\  
Adder (64)       &        8 &       67 &       10 & 190 \\  
Multiplier (13)  &        8 &   9 &        8 & 13 \\  
Multiplier (75)  &      486 & 674 &      597 &      873 \\  
QFT (18)         &        9 & 26 &       14 &       13 \\  
QFT (29)         &       14 & 48 &       24 &       20 \\  
QV (100)         &     6492 & 4237 &     6570 & 7554 \\  
\bottomrule
\end{tabular}
\end{table}

\begin{table}[H]
\centering
\caption{Distribution cost (EPR pairs consumed) on a 2-QPU network with nearest-neighbor intra-QPU connectivity.}
\label{tab:epr-2qpu-nearest-neighbor}
\begin{tabular}{lcccc}
\toprule
\textbf{Circuit (qubits)} & \textbf{Hypergraph} & \textbf{Dynamic Interaction} & \textbf{Static Interaction} & \textbf{Best Benchmark} \\
\midrule
Adder (28)       &       11 & 33 &       13 &       87 \\  
Adder (64)       &       11 & 92 &       13 & 206 \\  
Multiplier (13)  &       14 &  18 &       11 &       14 \\  
Multiplier (75)  &      679 & 980 &      638 &     1080 \\  
QFT (18)         &      162 & 141 &      162 &      162 \\  
QFT (29)         &      420 & 369 &      420 &      420 \\  
QV (100)         &     6492 & 4358 &     6474 & 7476 \\  
\bottomrule
\end{tabular}
\end{table}

\begin{table}[H]
\centering
\caption{Distribution cost (EPR pairs consumed) on a 3-QPU ring network, with nearest-neighbor intra-QPU connectivity.}
\label{tab:epr-ring-3qpu}
\begin{tabular}{lcccc}
\toprule
\textbf{Circuit (qubits)} & \textbf{Hypergraph} & \textbf{Dynamic Interaction} & \textbf{Static Interaction} & \textbf{Best Benchmark} \\
\midrule
Adder (28)       &        2 & 67 &       23 & 99 \\  
Adder (64)       &       12 & 144 &       23 & 291 \\  
Multiplier (13)  &       18 &  17 &       14 &       22 \\  
Multiplier (75)  &     1042 & 1828 &     1272 &     1215 \\  
QFT (18)         &      216 & 193 &      214 & 214 \\  
QFT (29)         &      552 & 459 & 527 &      528 \\  
QV (100)         &     8838 & 6738 &     8850 & 10137 \\  
\bottomrule
\end{tabular}
\end{table}

\begin{table}[H]
\centering
\caption{Distribution cost (EPR pairs consumed) on a 4-QPU grid network, with nearest-neighbor intra-QPU connectivity.}
\label{tab:epr-grid-4qpu}
\begin{tabular}{lcccc}
\toprule
\textbf{Circuit (qubits)} & \textbf{Hypergraph} & \textbf{Dynamic Interaction} & \textbf{Static Interaction} & \textbf{Best Benchmark} \\
\midrule
Adder (28)       &       33 & 117 & 83 & 153 \\  
Adder (64)       &       55 & 278 & 97 & 370 \\  
Multiplier (13)  &       27 &  39 & 27 &       34 \\  
Multiplier (75)  & 2739 & 3859 & 4022 & 3527 \\  
QFT (18)         & 226 & 205 & 206 & 190 \\  
QFT (29)         & 558 & 764 & 742 & 614 \\  
QV (100)         & 13651 & 18785 & 13488 & 13663 \\  
\bottomrule
\end{tabular}
\end{table}

\clearpage
\section{Segment-Length Sensitivity}
\label{app:segment-sensitivity}

The segment length $\ell$ was evaluated at fourteen values from $1$ to
$100$ for Multiplier~(13), QFT~(18), and Adder~(28) on a 2-QPU
nearest-neighbor network and a 4-QPU grid. Each data point corresponds to one
deterministic compilation.
Across the six circuit--network cases, the worst tested value cost
$1.47\times$ to $7.00\times$ the minimum (median $1.65\times$), while the
default from Eq.~\eqref{eq:segment-length} cost $1.05\times$ to $2.36\times$
the minimum (median $1.43\times$). All results in this paper use this default
without per-circuit tuning.

\begin{figure}[H]
  \centering
  \includegraphics[width=\linewidth]{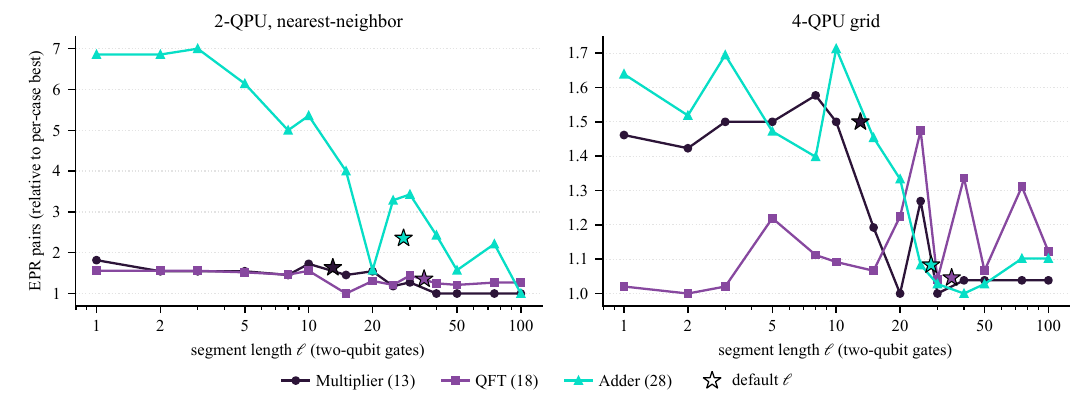}
  \caption{\textbf{EPR-pair cost as a function of segment length $\ell$}
  for three circuits on a 2-QPU nearest-neighbor network (left) and a 4-QPU
  grid (right). Each curve is normalized to its minimum over the sweep; stars
  mark the default from Eq.~\eqref{eq:segment-length}. Note the differing
  vertical scales.}
  \label{fig:segment-sensitivity}
\end{figure}

\clearpage
\section{Fan-Out Contention Circuit}

\begin{figure}[H]
  \centering
  \includegraphics[height=0.72\textheight,keepaspectratio]{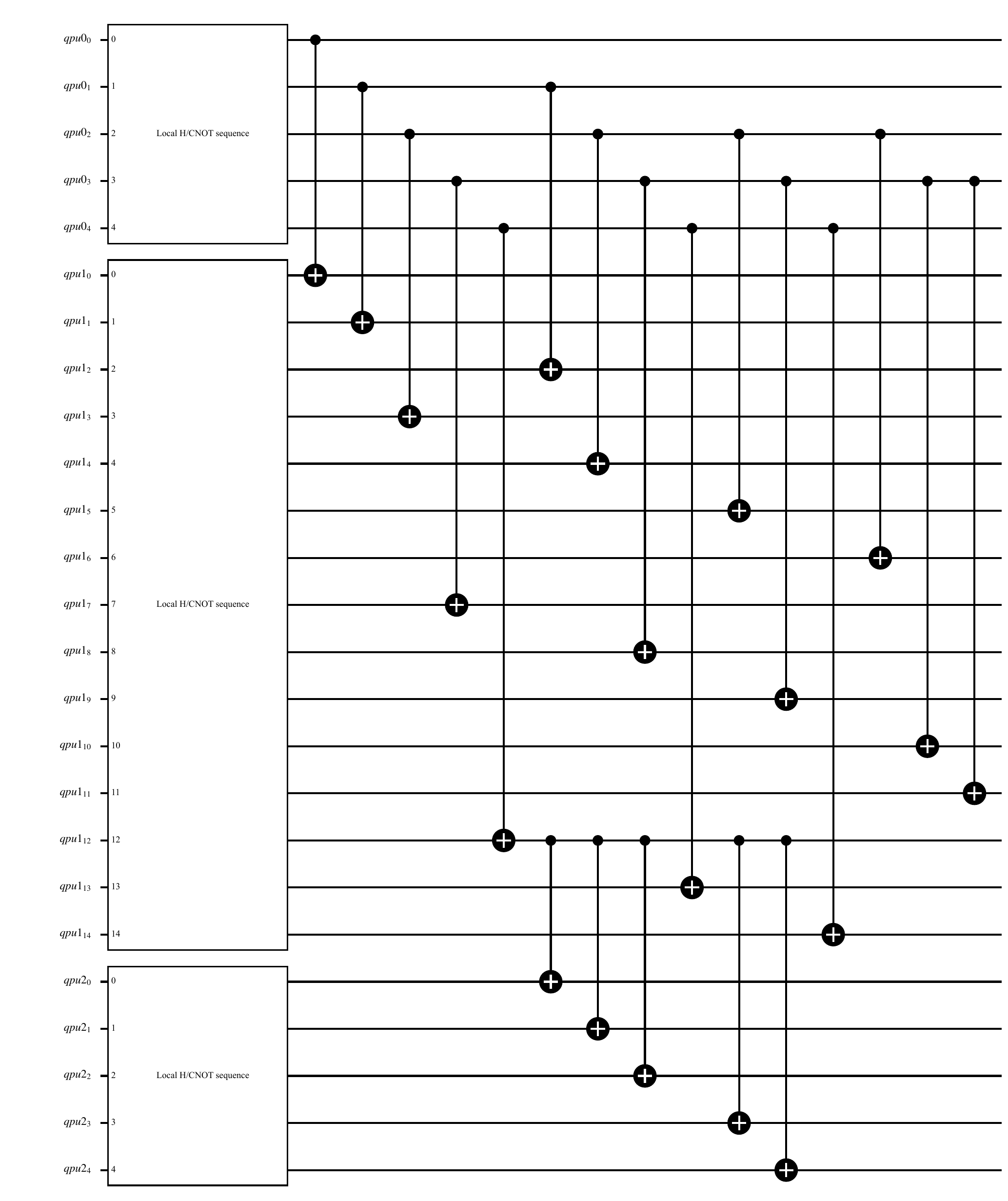}
  \caption{Qiskit rendering of the 3-QPU, 25-qubit fan-out circuit used in the scheduler experiment. For legibility, each QPU's sequence of local H gates and all-pairs local CNOTs is collapsed into a single labeled block; the inter-QPU CNOTs forming the upstream fan-outs and dependent downstream fan-out are shown explicitly.}
  \label{fig:fanout-contention-circuit}
\end{figure}

Figure~\ref{fig:fanout-contention-circuit} shows the circuit used in the controlled link-arbitration experiment. The three registers correspond to the three QPUs and contain 5, 15, and 5 computation qubits, respectively. Within each register, the local preparation sequence applies a Hadamard (H) gate to every qubit and then applies CNOT gates between every pair of local qubits.

\end{document}